\documentclass[10pt]{article}
\usepackage{amsmath,amssymb,amsbsy,graphicx, setspace,cite}
\usepackage{breqn,color}
\title{The tuned bistable nonlinear energy sink}
\author{Giuseppe Habib$^{1}$, Francesco Romeo$^{2}$}
\date{ }
\begin{document}
\maketitle
%\bigskip
%\date{
\center
{\em $^{1}$Department of Aerospace and Mechanical Engineering, University of Liege, Liege, Belgium. \\E-mail: giuseppe.habib@uniroma1.it}\\
%}
%Space Structures and Systems Laboratory, 
%\bigskip
{\em $^{2}$Department of Structural Engineering, Sapienza University of Rome, Rome, Italy
\\ E-mail: francesco.romeo@uniroma1.it} \\
%$^{2}$Department of Applied Mechanics, Budapest University of Technology and Economics, H-1521 Budapest, Hungary
%\\ E-mail: stepan@mm.bme.hu
%}}

%\doublespacing

%\abstract{
%}
\abstract{
A bistable nonlinear energy sink conceived to mitigate the vibrations of host structural systems is considered in this paper.
The hosting structure consists of two coupled symmetric linear oscillators (LOs) and the nonlinear energy sink (NES) is connected to one of them. The peculiar nonlinear dynamics of the resulting three-degree-of-freedom system is analytically described by means of its slow invariant manifold derived from a suitable rescaling, coupled with a harmonic balance procedure, applied to the governing equations transformed in modal coordinates.
On the basis of the first-order reduced model, the absorber is tuned and optimized to mitigate both modes for a broad range of impulsive load magnitudes applied to the LOs.
On the one hand, for low-amplitude, in-well, oscillations, the parameters governing the bistable NES are tuned in order to make it functioning as a linear tuned mass damper (TMD); on the other, for high-amplitude, cross-well, oscillations, the absorber is optimized on the basis of the invariant manifolds features.
The analytically predicted performance of the resulting tuned bistable nonlinear energy sink (TBNES) are numerically validated in terms of dissipation time; the absorption capabilities are eventually compared with either a TMD and a purely cubic NES.
It is shown that, for a wide range of impulse amplitudes, the TBNES allows the most efficient absorption even for the detuned mode, where a single TMD cannot be effective.}

\vspace{10pt}
\noindent Article published in Nonlinear Dynamics, 89(1), 179-196 (2017)
%\href{https://link.springer.com/article/10.1007/s11071-017-3444-y}{Nonlinear Dynamics, 89(1), 179-196 (2017)}

\section{Introduction}
Linear vibration absorbers represent a well established benchmark for mitigation of resonances, widely used in engineering practice with excellent performance. In this realm, tuned mass damper (TMD) devices have been extensively studied. Starting from the pioneering works \cite{Frahm, Denhartog} a rather extensive literature has been devoted to TMDs optimal design. Most of these studies have dealt with structural systems subjected to either harmonic or white noise excitations \cite{Warburton, Fujino,Rana}.
As known, in order to work properly, the TMD natural frequency must be tuned in the vicinity of the frequency of the resonance to be mitigated.
This implies that a single vibration absorber can be used to damp only one resonance of the primary structure.
As far as impulsive excitations, the vibration suppression of either a single-degree-of-freedom system (SDOF) and the dominant mode of a multiple-degree-of-freedom system (MDOF) was tackled in \cite{Salvi} by proposing an hybrid TMD, composed of an optimized TMD and a feedback closed-loop active controller.
Moreover, aiming to passively mitigate more than one mode in MDOF systems, the use of multiple TMDs has been also investigated \cite{chen2003}.
To overcome the TMD narrow frequency-band capabilities while relying on passive mitigation strategies based on a single device, many researchers studied the effect of additional nonlinearities in the absorber, aiming at letting the absorber resonate at more than one frequency.
This brought the development of the nonlinear energy sink (NES) consisting of a small mass connected to the primary system by an essential nonlinearity.
A number of NES designs were so far proposed, such as a cubic nonlinearity \cite{gendelman2001, kerschen2005}, vibro-impact device \cite{Vakakis2007impact}, eccentric rotator \cite{Sigalov2012} and tuned pendulum \cite{gendelman2015}.
More recently, the bistable NES (BNES), consisting of a small mass connected to the primary system by a spring with both cubic nonlinear and negative linear components \cite{Mikhlin,romeo2013,AlShudeifat} was also proposed.
While under impulsive excitations the NES dissipation mechanism hinges on the 1:1 transient resonance between the primary system and the NES, the so-called target energy transfer (TET) \cite{gendelman2001}, the BNES takes the additional advantage of chaotic intra-well motions \cite{romeo2015b}.
Both of them are not devoid of drawbacks. On the one hand the NES becomes almost ineffective below a certain energy threshold; on the other hand once the BNES dynamics is limited to in-well oscillations it becomes practically unable to dissipate arbitrary small oscillations.

In this paper we propose a variation of the classical BNES consisting in tuning the natural frequency of its in-well dynamics to one of the frequencies of the primary system, in the same fashion of the classical linear tuned mass damper (TMD).
Moreover, based on the procedure presented in \cite{Zulli,Luongo2013,Zulli2015}, we provide an analytical framework for guiding the BNES parameters optimization with respect to the high-amplitude cross-well oscillations. 
This tuned BNES (TBNES) allows to increase the operational range of the BNES to small oscillation amplitudes, while maximizing its remarkable performance for intermediate and high amplitudes.
Acknowledging the difference between the two, the TBNES reminds the nonlinear tuned vibration absorber studied in \cite{habib2015,habibPRSA,habib2016}.
In fact, both of them exploit effective linear dynamics at low amplitude, while they take advantage of nonlinearities for large oscillation amplitudes.

The paper is organized as follows.
At first, in Section 2, the governing equations and their transformation in modal coordinates are introduced, the absorber tuning based on the in-well dynamics is then described.
Then, in Section 3, the TBNES parameters are optimized, according to the system slow invariant manifold.
After a numerical validation of the proposed design (Section 4), in Section 5 a comparison among the TBNES, the NES and the TMD performance is computed. 

\section{Governing equations and absorber tuning}
\begin{figure}
\begin{centering}
\includegraphics[width =0.6\textwidth]{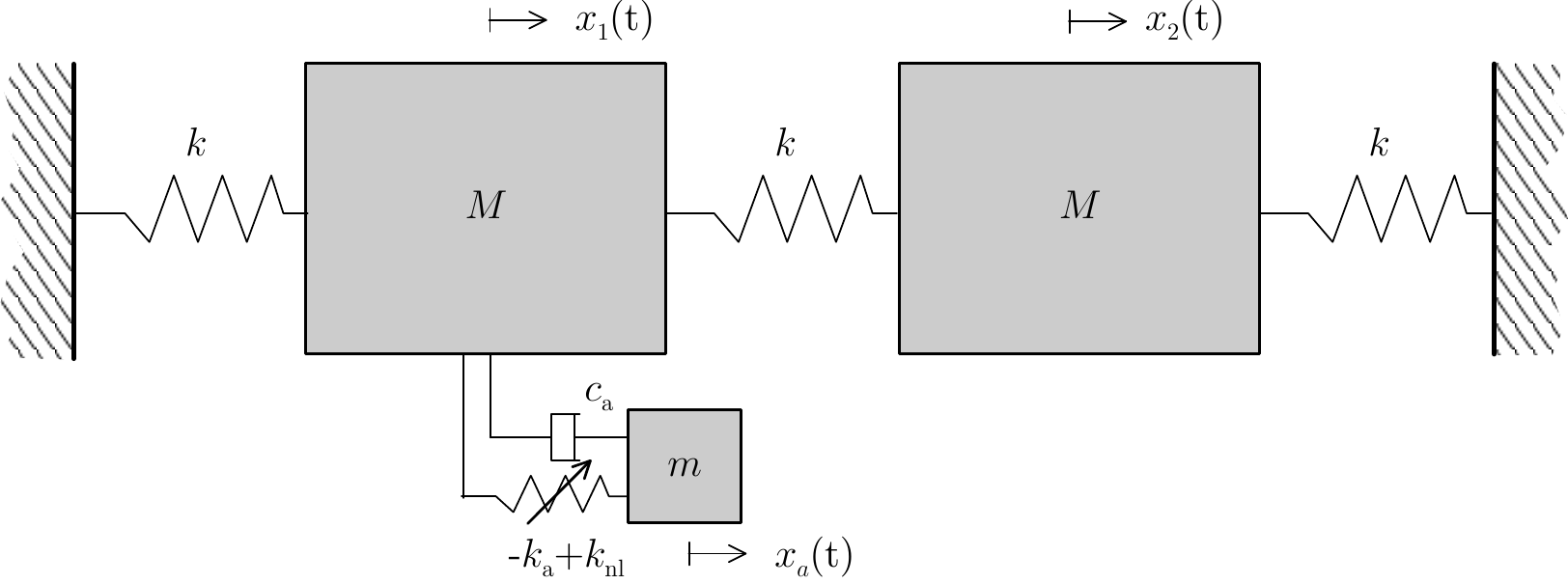}
\par\end{centering}
\caption{\label{model} A three-degree-of-freedom system consisting of two coupled symmetric linear oscillators (LOs) and a bi-stable absorber connected to one of them.}
\end{figure}
We consider a simple model shown in Fig.~\ref{model}. The dynamics of this system is governed by the equations 
\begin{equation}
\begin{split}
&Mx_1''+kx_1+k\left(x_1-x_2\right)-k_a\left(x_1-x_a\right)+k_{nl}\left(x_1-x_a\right)^3+c_a\left(x_1'-x_a'\right)=0\\
&Mx_2''+kx_2+k\left(x_2-x_1\right)=0\\
&mx_a''-k_a\left(x_a-x_1\right)+k_{nl}\left(x_a-x_1\right)^3+c_a\left(x_a'-x_1'\right)=0
\end{split}
\label{first}
\end{equation}
where $x_1$ and $x_2$ refer to the displacements of the primary 2 DoF system, while $x_a$ refers to the displacement of the bistable absorber; $m$ is assumed much smaller than $M$ and the prime denote differentiation with respect to time $t$.
The choice of considering an undamped hosting structure reflects results obtained in previous works \cite{romeo2015dynamics}, where it is illustrated that small damping in the primary system does not affect the overall qualitative dynamics.

Dividing the system of equations (\ref{first}) by $M$, introducing the dimensionless time $T=\omega_nt$, where $\omega_n=\sqrt{k/M}$, and then dividing the system by $\omega_n^2$ we obtain \begin{equation}\begin{split}
&\ddot x_1+2x_1-x_2=\gamma^2\varepsilon\left(x_1-x_a\right)-\lambda_3\varepsilon\left(x_1-x_a\right)^3-2\mu_2\varepsilon\left(\dot x_1-\dot x_a\right)\\
&\ddot x_2+2x_2-x_1=0\\
&\varepsilon\left(\ddot x_a-\gamma^2\left(x_a-x_1\right)+\lambda_3\left(x_a-x_1\right)^3+2\mu_2\left(\dot x_a-\dot x_1\right)\right)=0,
\end{split}\label{reducedinx}\end{equation}
where $\gamma=\omega_a/\omega_n=\sqrt{\left(k_a/m\right)/\left(k/M\right)}$, $\lambda_3=k_{nl}/\left(m\omega_n^2\right)$, $\varepsilon=m/M$, $\mu_2=c_a/\left(2m\omega_n\right)$, $\Omega=\omega/\omega_n$ and the overdots denote differentiation with respect to $T$.

Introducing the variables $y_1=\left(x_1+x_2\right)/2$, $y_2=\left(x_1-x_2\right)/2$ and $y_3=x_1-x_a$, the governing equations (\ref{reducedinx}) are transformed in the primary system modal coordinates, i.e. \begin{equation}
\begin{split}
&\ddot y_1+y_1=\frac{1}{2}\gamma^2\varepsilon y_3-\frac{1}{2}\varepsilon\lambda_3y_3^3-\mu_2\varepsilon\dot y_3\\
&\ddot y_2+3y_2=\frac{1}{2}\gamma^2\varepsilon y_3-\frac{1}{2}\varepsilon\lambda_3y_3^3-\mu_2\varepsilon\dot y_3\\
&\ddot y_3+y_1+3y_2+\left(1+\varepsilon\right)\left(-\gamma^2y_3+\lambda_3y_3^3+2\mu_2\dot y_3\right)=0
\end{split}\label{reducediny}
\end{equation}
Starting from equations (\ref{reducediny}), by considering $\varepsilon\ll1$ as a perturbation parameter, an analytical framework enabling to design the bistable absorber and to optimize its performance with respect to the different dynamic regimes experienced by the system is derived. In particular, for a small, fixed value of mass ratio $\varepsilon$, the role played by the frequency ratio $\gamma$, the cubic stiffness parameter $\lambda_3$ and the damping parameter $\mu_2$ will be thoroughly discussed.

As pointed out in \cite{Sigalov2012}, where a similar system as considered, the choice of such a simple primary system is not restrictive. The underlying assumptions are that the two natural frequencies of the linear oscillators are of the same order of magnitude, incommensurate and remote.
Thus, the special case of internal resonance  is not here considered.

Assuming $y_1$ and $y_2$ periodic, the third equation of (\ref{reducediny}) corresponds to a quasiperiodically excited, linearly damped, Duffing oscillator with a negative linear restoring force.
The dynamics of such a system, which cannot be solved analytically, has been extensively studied in the literature  \cite{Guck, Kovacic} and can lead to various dynamical phenomena, including chaos.
Three main different scenarios are considered: periodic (or quasiperiodic if both modes are excited) in-well motions, chaotic cross-well motions and periodic (or quasiperiodic) large cross-well motions.
In the latter case, the cubic stiffness term is dominant over the negative linear one, thus the system is not largely affected by the two potential wells and the solution remains symmetric.
Since this motion involves large oscillation amplitudes, a different analytical framework will be developed for its analysis in Section 3.
In the other cases, the two potential wells dominate the dynamics, which is mainly affected by the energy necessary to pass from one well to the other, rather than by the shape of the restoring force.
In the following, tuning and optimization of the NES parameters are targeted to exploit these three dynamic regimes for maximizing the energy taken out of the primary system.

\begin{figure}
\begin{centering}
\includegraphics[trim = 15mm 5mm 5mm 15mm,width=0.4\textwidth]{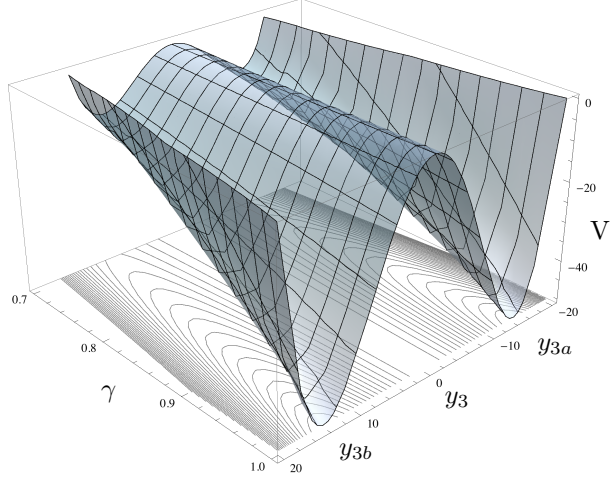}
\put(-50.,20){$y_{3}$}
\put(-15.,40){$y_{3a}$}
\put(-90.,0){$y_{3b}$}
\put(-180.,22){$\gamma$}
\put(5,82){V}
\par\end{centering}
\caption{\label{potential}Potential energy of the absorber; $\frac{1}{\sqrt{2}} \leqslant \gamma \leqslant 1$, $\lambda_3=0.005$.}
\end{figure}

By considering at first the in-well dynamics, the potential energy $V$ of the absorber, represented in Fig.~\ref{potential} as a function of $\gamma$, shows two stable equilibria for $y_{3a}=-\gamma/\sqrt{\lambda_3}$ and $y_{3b}=\gamma/\sqrt{\lambda_3}$ and one unstable equilibrium at the origin.
Centering Eqs.~(\ref{reducediny}) around one of the stable equilibria ($\tilde y_{3}=y_{3}-\gamma/\sqrt{\lambda_3}$), we obtain the system of equations \begin{equation}
\begin{split}
&\ddot y_1+y_1+\varepsilon\left(\gamma  \mu_2 \dot{\tilde y}_3+\gamma ^2 \tilde y_3+\frac{3}{2} \gamma  \sqrt{\lambda_3} \tilde y_3^2 +\frac{1}{2} \lambda_3 \tilde y_3^3 \right)=0\\
&\ddot y_2+3y_2+\varepsilon\left(\gamma  \mu_2 \dot{\tilde y}_3+\gamma ^2 \tilde y_3+\frac{3}{2} \gamma  \sqrt{\lambda_3} \tilde y_3^2 +\frac{1}{2} \lambda_3 \tilde y_3^3 \right)=0\\
&\ddot{\tilde y}_3+y_1+3 y_2+\left(1+\varepsilon\right)\left(2 \gamma  \mu_2 \dot{\tilde y}_3+2 \gamma ^2 \tilde y_3+3 \gamma  \sqrt{\lambda_3} \tilde y_3^2  +\lambda_3 \tilde y_3^3 \right)=0.
\end{split}\label{centred}
\end{equation}
For small values of $y_1$ and $y_2$, the absorber oscillates around one of its stable positions. In order to go from one potential well to the other one, an energy level of at least $\gamma^4/\left(4\lambda_3\right)$ (with respect to the energy of the system at rest in a stable equilibrium) must be reached.

In-well motions involve small values of $y_1$, $y_2$ and $y_3$, which are thus assumed of order $\varepsilon$.
Neglecting terms of higher order in $\varepsilon$ in Eqs.~(\ref{centred}), we attain the linear system
\begin{equation}
\begin{split}
&\ddot y_1+y_1=0\\
&\ddot y_2+3y_2=0\\
&\ddot{\tilde y}_3+y_1+3 y_2+2 \gamma  \mu_2 \dot{\tilde y}_3+2 \gamma ^2 \tilde y_3=0,
\end{split}\label{linearized}
\end{equation}
which corresponds to a 2 DoF primary system with an attached TMD.
In order to enforce a 1:1 resonance between the primary system and the absorber, $\gamma$ should be tuned either at $\gamma=1/\sqrt{2}$, for the first mode of vibration, or at $\gamma=\sqrt{3/2}$ for the second mode.

If nonlinear terms are not neglected, as exhaustively explained in \cite{Kovacic, Nayfeh}, the in-well periodic oscillations undergo a softening effect.
The frequency backbone of the in-well vibrations is described by the equation $\omega_n=\sqrt 2\gamma-\left(Y_{3}/0.9710\right)^2\lambda_3/\gamma$, where $Y_{3}$ is the amplitude of oscillation.
This suggests that, if $\gamma$ is chosen to meet a perfect 1:1 resonance at low oscillation amplitude, as the latter grows the absorber detunes. Indeed, a higher value of $\gamma$ would be required to counteract the bistable absorber softening in order to keep perfect resonance condition; the ensuing potential energy surface is shown in Fig.~\ref{potential} within the response amplitude range of interest.
However, as it will be numerically evidenced in Section \ref{numerical_validation}, a single choice of tuning, based on the low amplitude oscillations frequency, guarantees efficient vibration absorption for the whole in-well dynamics.
If a sufficient energy level is reached by the system, the absorber undergoes chaotic behavior, the onset of which, as a rule,  is facilitated by small values of damping $\mu_2$ and $\gamma$, and large values of $\lambda_3$. The description of such dynamics is extensively studied in literature (e.g. \cite{Guck}); more specifically, as far as the energy dissipation is concerned, the appearance of chaotic dynamics allows to enlarge the bandwidth of effectiveness of the absorber, as it was recently shown for a single-degree-of-freedom primary system in \cite{romeo2015b}.
The detailed description of this dynamic regime is out of the scope of this paper; anyway it is worth anticipating that, as confirmed by the numerical simulations reported in Section \ref{numerical_validation}, large amplitude, cross-well periodic motions are relatively more efficient, in terms of energy dissipation time, than chaotic dynamics.

\section{Analytical optimization of the tuned absorber}

In order to investigate large periodic motions, a different scaling of the system parameters is adopted.
$y_1$, $y_2$ and $y_3$ are assumed of order $\varepsilon^{-1}$, while $\lambda_3$ of order $\varepsilon^2$.
Collecting then terms of order $\varepsilon^{-1}$, Eqs.~(\ref{reducediny}) are reduced to \begin{eqnarray}
&&\ddot y_1+y_1=0\label{order-1a}\\
&&\ddot y_2+3y_2=0\label{order-1b}\\
&&\ddot y_3+2\mu_2 \dot y_3-\gamma^2y_3+\lambda_3 y_3^3=-y_1-3y_2.
\label{order-1c}
\end{eqnarray}
The motions relative to Eqs.~(\ref{order-1a})-(\ref{order-1c}) are self symmetric and, unlike the in-well motions, they have no symmetric counterpart.
Therefore, it is not convenient to center the system around on of the stable equilibria.

In order to define an approximate solution, we adopt the harmonic balance method \cite{Zulli,Luongo2013}, by assuming 1:1 resonance between the primary system and the absorber.
The solutions of Eqs.~(\ref{order-1a}) and (\ref{order-1b}) are \begin{equation}
y_1=A_1e^{iT}+\text{c.c.}\qquad\text{and}\qquad y_2=A_2e^{\sqrt 3iT}+\text{c.c.},
\end{equation}
where $A_1$ and $A_2$ are complex and c.c. stands for complex conjugate.
The approximate solution of Eq.~(\ref{order-1c}) is expressed by \begin{equation}
y_3=B_1\left(t_1\right)e^{iT}+B_2\left(t_1\right)e^{\sqrt 3iT}+\text{c.c.}\label{approx}
\end{equation}
We substitute Eq.~(\ref{approx}) into Eq.~(\ref{order-1c}) and collect harmonics of $e^{iT}$ and $e^{\sqrt 3iT}$, obtaining \begin{equation}\begin{split}
\left(e^{it}\right):&\,-B_1+A_1-\gamma^2B_1+\lambda_3\left(3B_1^2 \bar B_1+6B_1\bar B_1 B_2\right)+2\mu_2 iB_1=0\\
\left(e^{\sqrt 3it}\right):&\,-3B_2+3A_2-\gamma^2B_2+\lambda_3\left(3B_2^2 \bar B_2+6B_1B_2 \bar B_2\right)+2\mu_2 iB_2=0.\end{split}\label{HB1}
\end{equation}
It is worth noticing that, since the adopted harmonic balance procedure does not take into account the stability of solutions, the destabilizing effect of the negative spring on these symmetric solutions is overlooked.
However, the negative spring is relevant for the stability only at low oscillation amplitudes, for which the considered scaling is not valid.

By defining $B_1=1/2b_1e^{i\beta_1}$, $B_2=1/2b_2e^{i\beta_2}$, $A_1=1/2a_1e^{i\alpha_1}$ and $A_2=1/2a_2e^{i\alpha_2}$, and separating real and imaginary parts of the first equation of (\ref{HB1}), we have \begin{equation}
\begin{split}
\frac{1}{2}a_1\cos\alpha_1&=\frac{1}{2}b_1\cos\beta_1\left(1+\gamma^2 -\frac{3}{4}\lambda_3b_1^2\right)-\frac{3}{4}\lambda_3b_1^2b_2\cos\beta_2+\mu_2b_1\sin\beta_1\\
\frac{1}{2}a_1\sin\alpha_1&=\frac{1}{2}b_1\sin\beta_1\left(1+\gamma^2 -\frac{3}{4}\lambda_3b_1^2\right)-\frac{3}{4}\lambda_3b_1^2b_2\sin\beta_2-\mu_2b_1\cos\beta_1.
\end{split}\label{imre1}
\end{equation}
We calculate the squares of the two equations of (\ref{imre1}) and we sum them up attaining  \begin{equation}
\begin{split}
&\frac{a_1^2}{4}-\frac{b_1^2}{4}\left(1+\gamma^2-\frac{3}{4}\lambda_3b_1^2\right)^2-\mu_2^2b_1^2-\frac{9}{16}\lambda_3^2b_1^4b_2^2=-\frac{3}{4}\lambda_3b_1^3b_2\left(1+\gamma^2-\frac{3}{4}\lambda_3b_1^2\right)\cos\left(\beta_1-\beta_2\right)\\
&\quad -\frac{3}{2}\mu_2\lambda_3b_1^3b_2\sin\left(\beta_1-\beta_2\right).
\end{split}\label{man1}
\end{equation}
Repeating the same operation with the second equation of (\ref{HB1}) we obtain \begin{equation}
\begin{split}
&\frac{9a_2^2}{4}-\frac{b_2^2}{4}\left(3+\gamma^2-\frac{3}{4}\lambda_3b_2^2\right)^2-\mu_2^2b_2^2-\frac{9}{16}\lambda_3^2b_1^2b_2^4=-\frac{3}{4}\lambda_3b_1b_2^3\left(3+\gamma^2-\frac{3}{4}\lambda_3b_2^2\right)\cos\left(\beta_1-\beta_2\right)\\
&\quad +\frac{3}{2}\mu_2\lambda_3b_1b_2^3\sin\left(\beta_1-\beta_2\right).
\end{split}\label{man2}
\end{equation}

Equations (\ref{man1}) and (\ref{man2}) describe the invariant manifold that relates the slow dynamics of $y_3$ with respect to $y_1$ and $y_2$; a detailed analysis of the obtained manifold is performed in the following by considering separately the cases in which the excitation involves either a single mode or both modes.

\subsection{Single mode dynamics}

We consider at first the case when only the first mode of the primary system is initially excited.
In this case $y_2$ is assumed of order $\varepsilon$, and, repeating the same analysis carried out in the previous section, we obtain the invariant manifold
\begin{equation}
a_1^2=b_1^2\left(\left(1+\gamma^2-\frac{3}{4}\lambda_3b_1^2\right)^2 +4\mu_2^2\right)
\label{many1}
\end{equation}
in which, having selected the first mode, $a_2$ and $b_2$ do not appear since the terms in $y_2$ vanish in the first order dynamics.

\begin{figure}
\begin{centering}
\setlength{\unitlength}{\textwidth}
\begin{picture}(1,1.2)
\put(0.07,0.80){\includegraphics[trim = 10mm 10mm 10mm 10mm,width=0.4\textwidth]{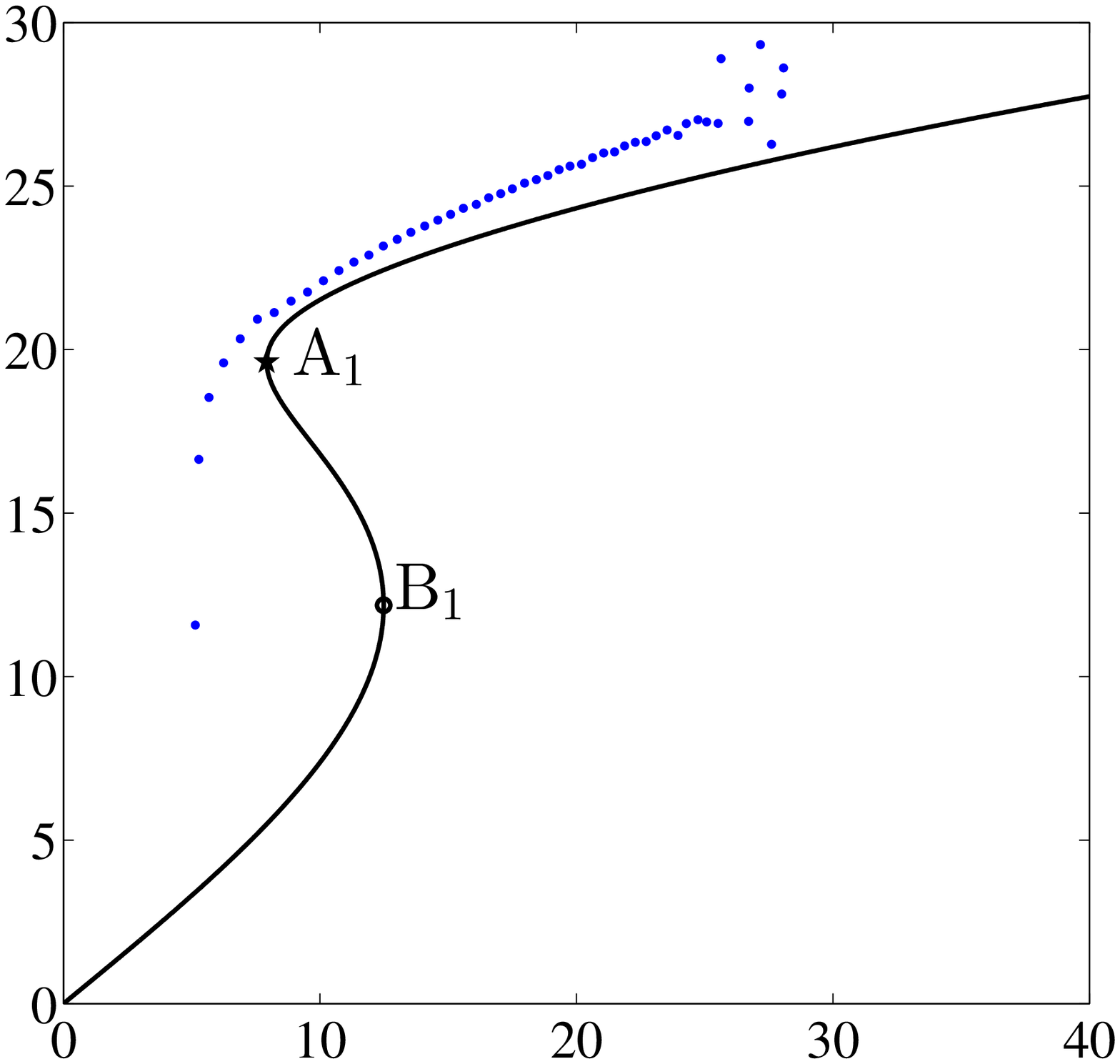}}
\put(0.51,0.80){\includegraphics[trim = 10mm 10mm 10mm 10mm,width=0.4\textwidth]{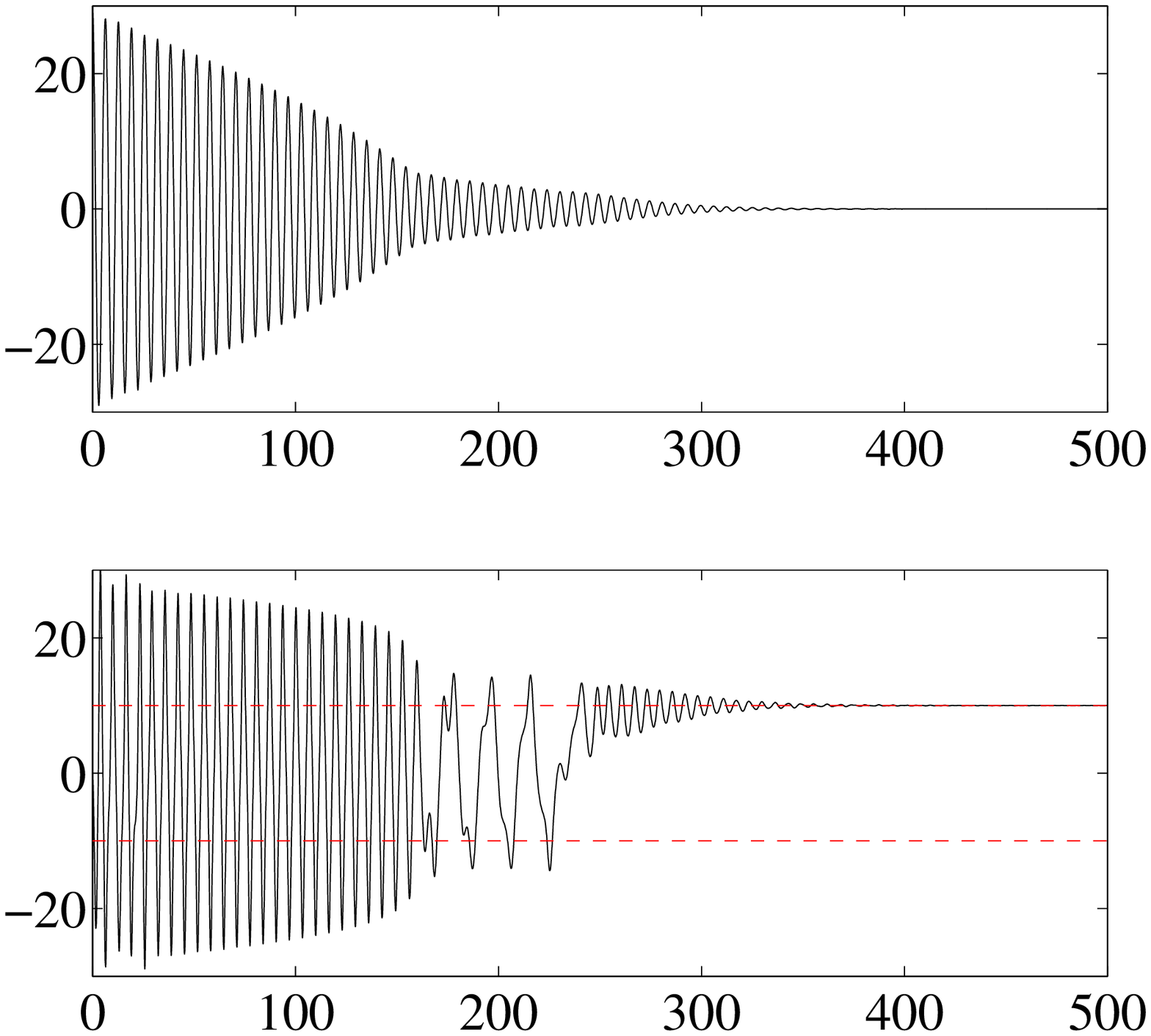}}
\put(0.07,0.4){\includegraphics[trim = 10mm 10mm 10mm 10mm,width=0.4\textwidth]{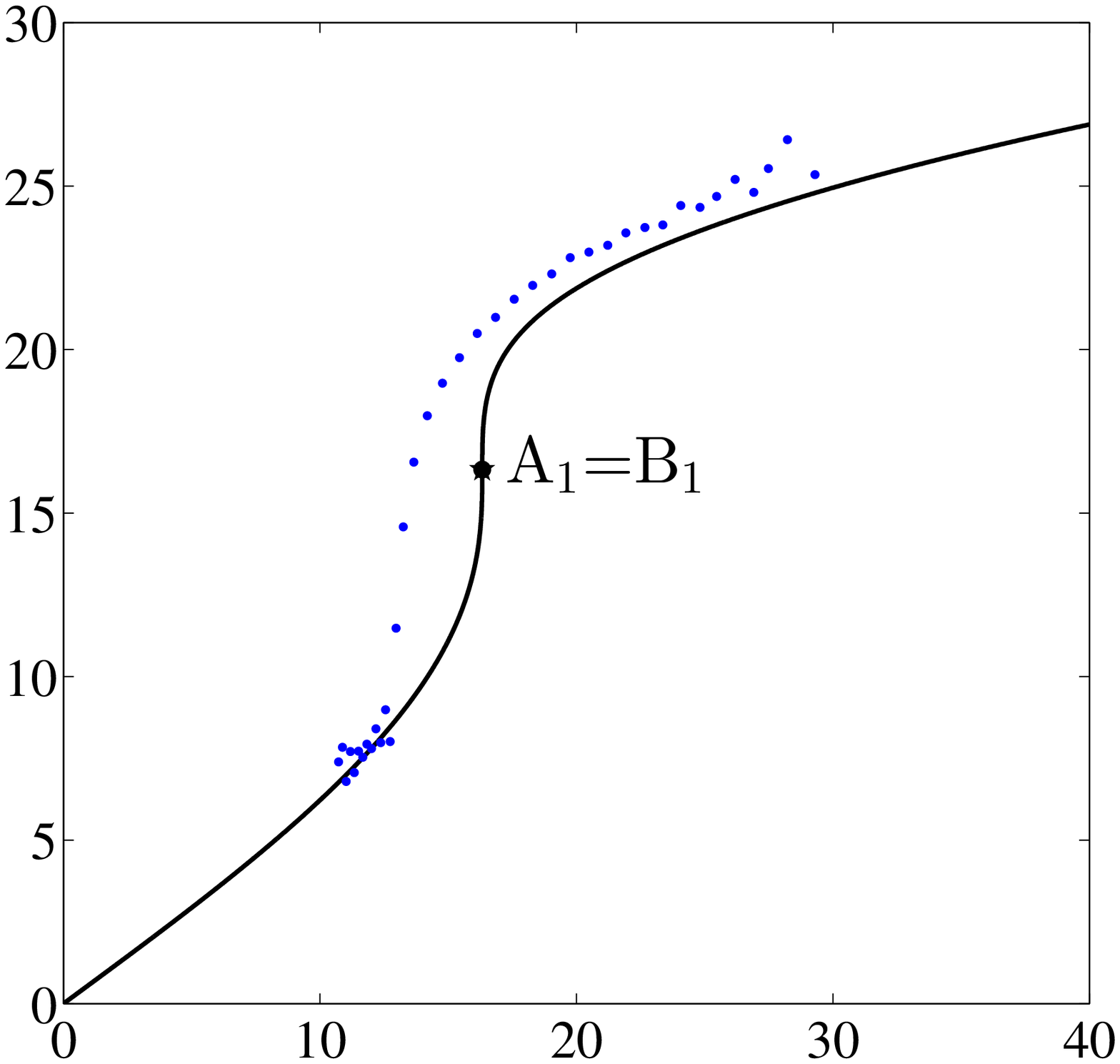}}
\put(0.51,0.4){\includegraphics[trim = 10mm 10mm 10mm 10mm,width=0.4\textwidth]{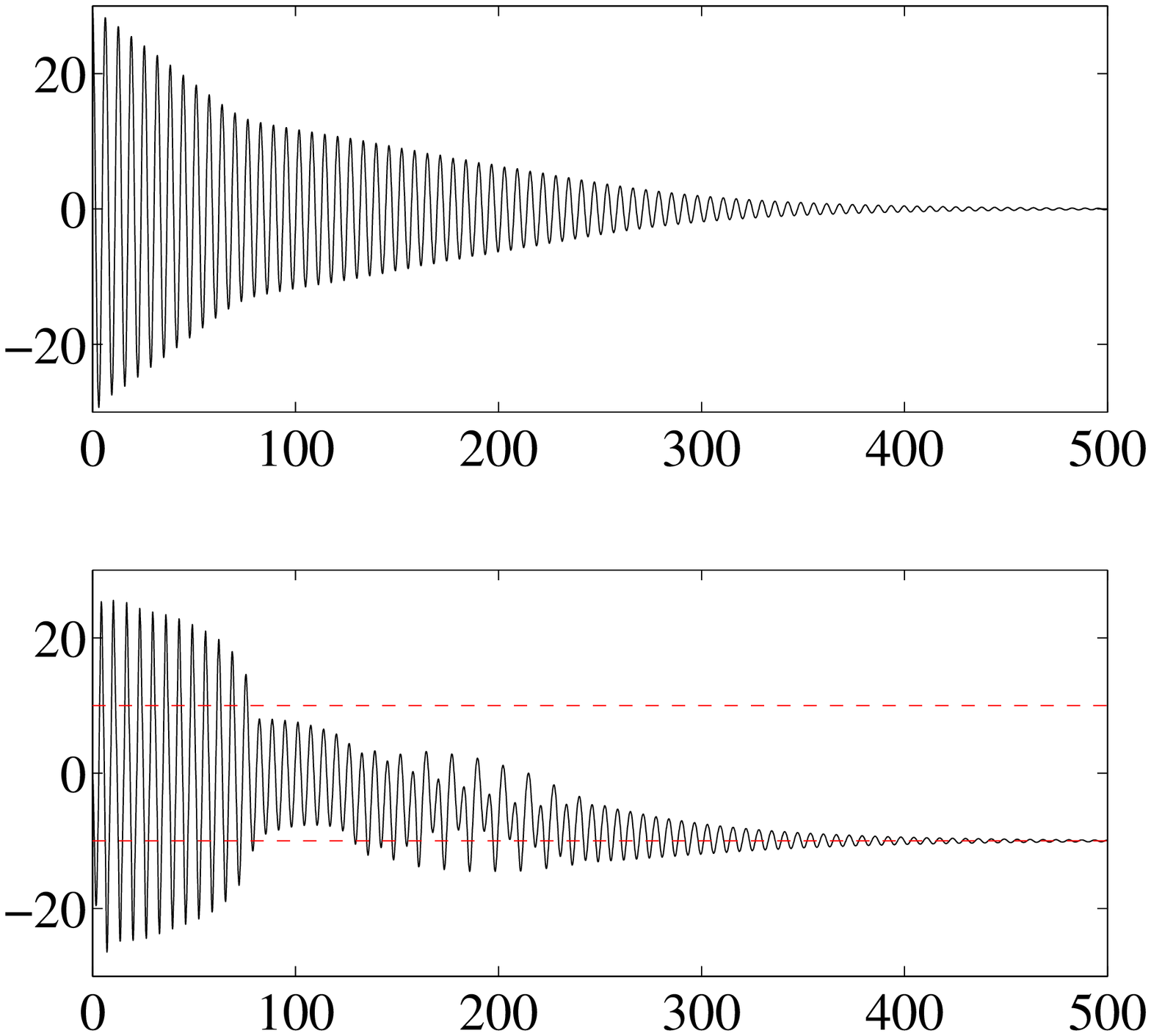}}
\put(0.07,0.0){\includegraphics[trim = 10mm 10mm 10mm 10mm,width=0.4\textwidth]{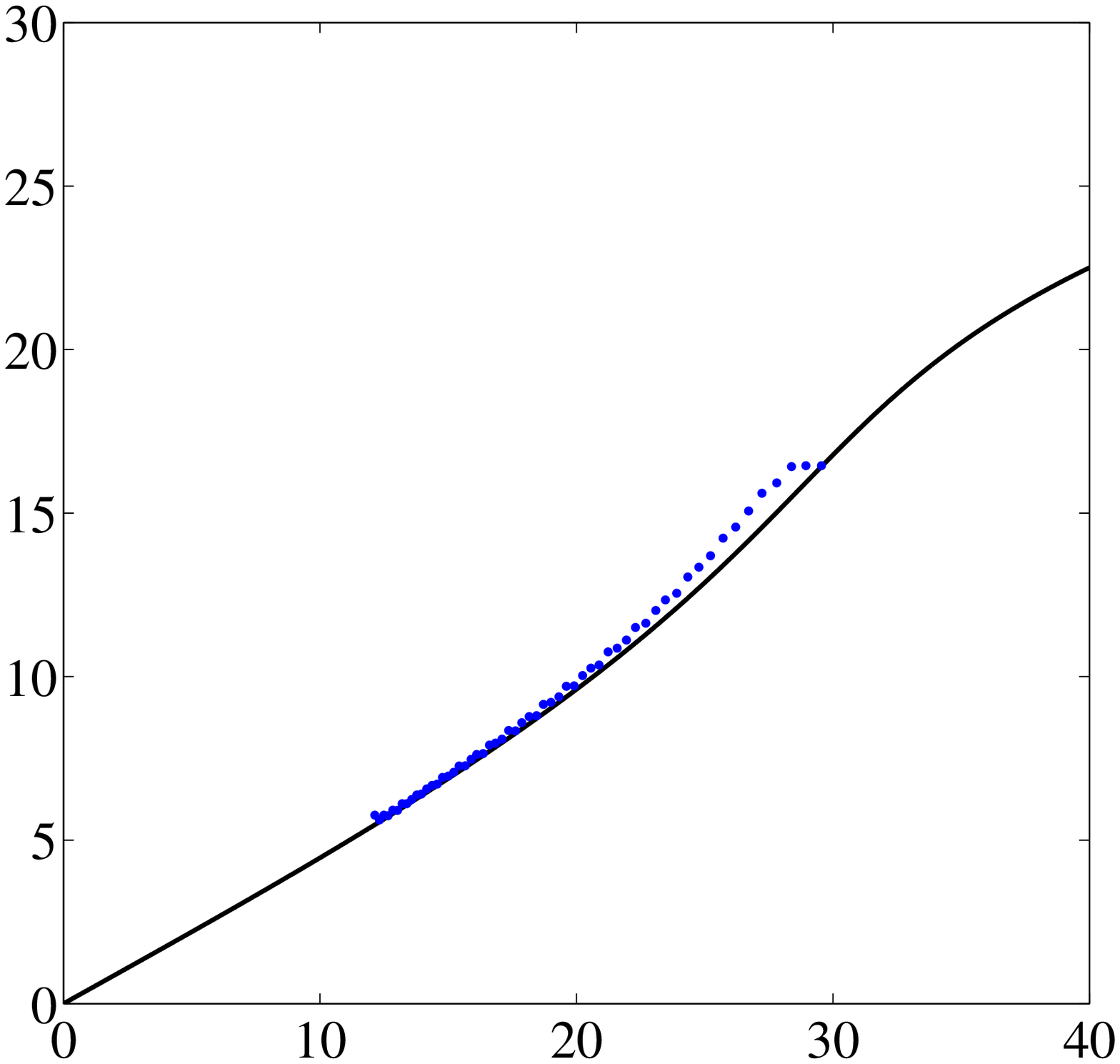}}
\put(0.51,0.0){\includegraphics[trim = 10mm 10mm 10mm 10mm,width=0.4\textwidth]{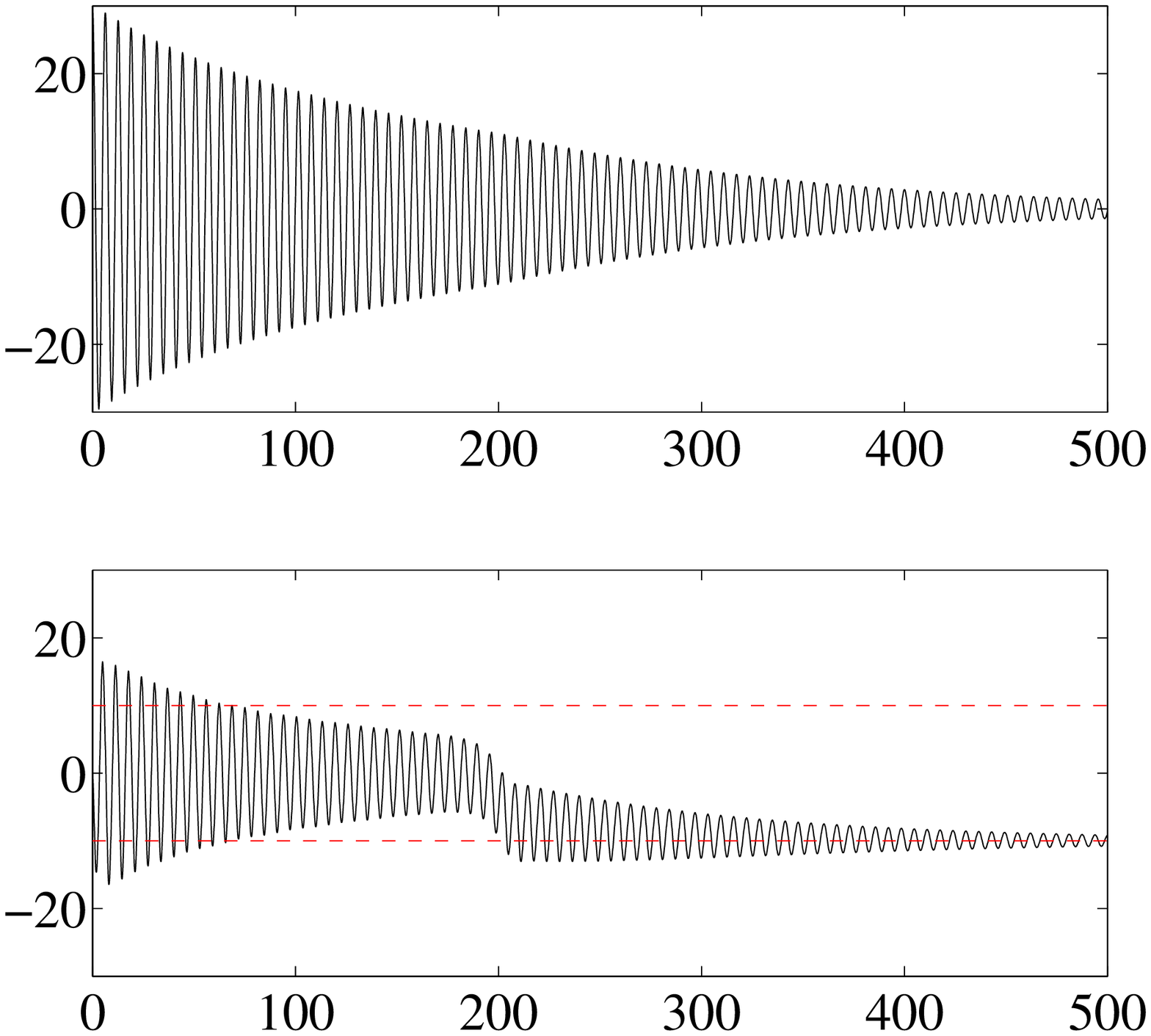}}
\put(0.05,1.13){(a)}
\put(0.49,1.13){(b)}
\put(0.05,0.73){(c)}
\put(0.49,0.73){(d)}
\put(0.05,0.33){(e)}
\put(0.49,0.33){(f)}
\put(0.06,0.19){$b_1$}
\put(0.06,0.59){$b_1$}
\put(0.06,0.99){$b_1$}
\put(0.5,0.28){$y_1$}
\put(0.5,0.09){$y_3$}
\put(0.5,0.68){$y_1$}
\put(0.5,0.49){$y_3$}
\put(0.5,1.08){$y_1$}
\put(0.5,0.89){$y_3$}
\put(0.275,-0.01){$a_1$}
\put(0.715,-0.01){$T$}
\put(0.275,0.39){$a_1$}
\put(0.715,0.39){$T$}
\put(0.275,0.79){$a_1$}
\put(0.715,0.79){$T$}
\par
\end{picture}
\end{centering}
\caption{\label{manifold}(a,c,e) Invariant manifolds obtained from Eq.~(\ref{many1}) for $\gamma=0707$, $\lambda_3=0.005$, $\mu_2=0.2$ (a), $\mu_2=\mu_2^\star=0.433$ (c) and $\mu_2=2\mu_2^\star=0.866$ (e). Blue dots depict the peaks identified from the time series shown in the subplots (b,d,f). (b,d,f) Time series for the same parameters utilized for the manifold calculation ($\varepsilon=0.05$).}
\end{figure}

Figure \ref{manifold}a illustrates the invariant manifold for $\gamma=0.707$, $\lambda_3=0.005$ and $\mu_2=0.2$.
The manifold displays the classical \emph{S} shape, which generates relaxation oscillations in the forced case [cite].
The relevance of the manifold for the description of the free dynamics of the system is proven by the blue dots, marking the peaks of the free decay, and overlapping the invariant manifold.
As expected, when the system reaches point A$_1$ of the manifold, there is a sudden decrease of the amplitude of oscillation of the absorber ($y_3$), which results in a deterioration of its absorption performance.

Explicit equations of the coordinates of points A$_1$ and B$_1$, marking the folding of the manifold, can be easily obtained and are given by
\begin{equation}
\begin{split}
\text{A}_1=&\left(\frac{\sqrt{4\left(2 \gamma ^2+2+\varphi\right) \left(\frac{1}{9} \left(\gamma ^2+1-\varphi\right)^2+4 \mu_2^2\right)}}{3 \sqrt{\lambda_3}},\frac{\sqrt{4 \left(2 \gamma ^2+2+\varphi\right)}}{3 \sqrt{\lambda_3}}\right)\\
\text{B}_1=&\left(\frac{\sqrt{4\left(2 \gamma ^2+2-\varphi\right) \left(\frac{1}{9} \left(\gamma ^2+1+\varphi\right)^2+4 \mu_2^2\right)}}{3 \sqrt{\lambda_3}},\frac{\sqrt{4\left(2 \gamma ^2+2-\varphi\right)}}{3 \sqrt{\lambda_3}}\right),
\end{split}\label{AB1}
\end{equation}
where $\varphi=\sqrt{\left(\gamma ^2+1\right)^2-12 \mu_2^2}$. Points A$_1$ and B$_1$ delimit a region of bistable behavior of the absorber. The nature of this bistability is not at all related to the double-well potential of the absorber. In fact, it exists also for the NES (imposing $\gamma=0$ the TBNES is reduced to an NES).
This bistability is instead related to the coexistence of a small and a large periodic response of the absorber.

A$_1$ and B$_1$ get closer to each other for increasing values of $\mu_2$, until they merge for $\mu_2=\mu_2^\star=\left(1+\gamma^2\right)/\left(2\sqrt 3\right)$, which corresponds to \begin{equation}
\text{A}_1=\text{B}_1=\left(\sqrt{\frac{32 \left(\gamma ^2+1\right)^3}{81 \lambda_3}},\sqrt{\frac{8 \left(\gamma ^2+1\right)}{9 \lambda_3}}\right).
\end{equation}
The invariant manifold for $\mu_2=\mu_2^\star=0.433$ is illustrated in Fig.~\ref{manifold}(c), with the corresponding time series of a free vibration decay.
Increasing even more the absorber damping $\mu_2>\mu_2^\star$, the manifold can be almost linearized, as shown in Fig.~\ref{manifold}(e) for $\mu_2=2\mu_2^\star=0.866$.

A qualitative comparison of the time series in Fig.~\ref{manifold} highlights that, as damping increases, chaotic dynamics, which separates the periodic cross-well motions and in-well motions, is damped out.
This fact, anyway expected, is not analyzed in further details.

\begin{figure}
\begin{centering}
\setlength{\unitlength}{\textwidth}
\begin{picture}(1,0.33)
\put(0.04,0.0){\includegraphics[trim = 10mm 10mm 10mm 10mm,width=0.4\textwidth]{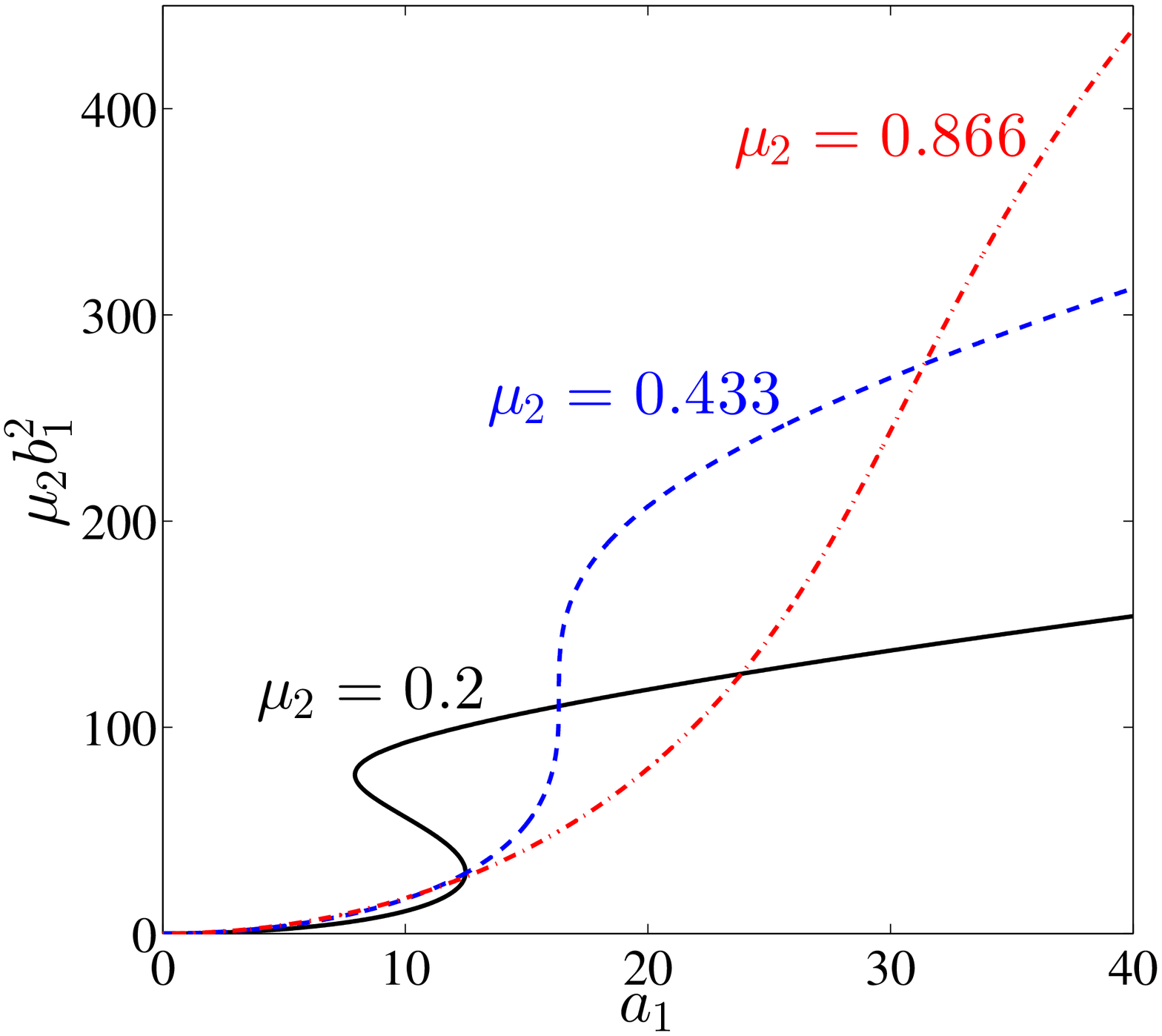}}
\put(0.48,0.0){\includegraphics[trim = 10mm 10mm 10mm 10mm,width=0.4\textwidth]{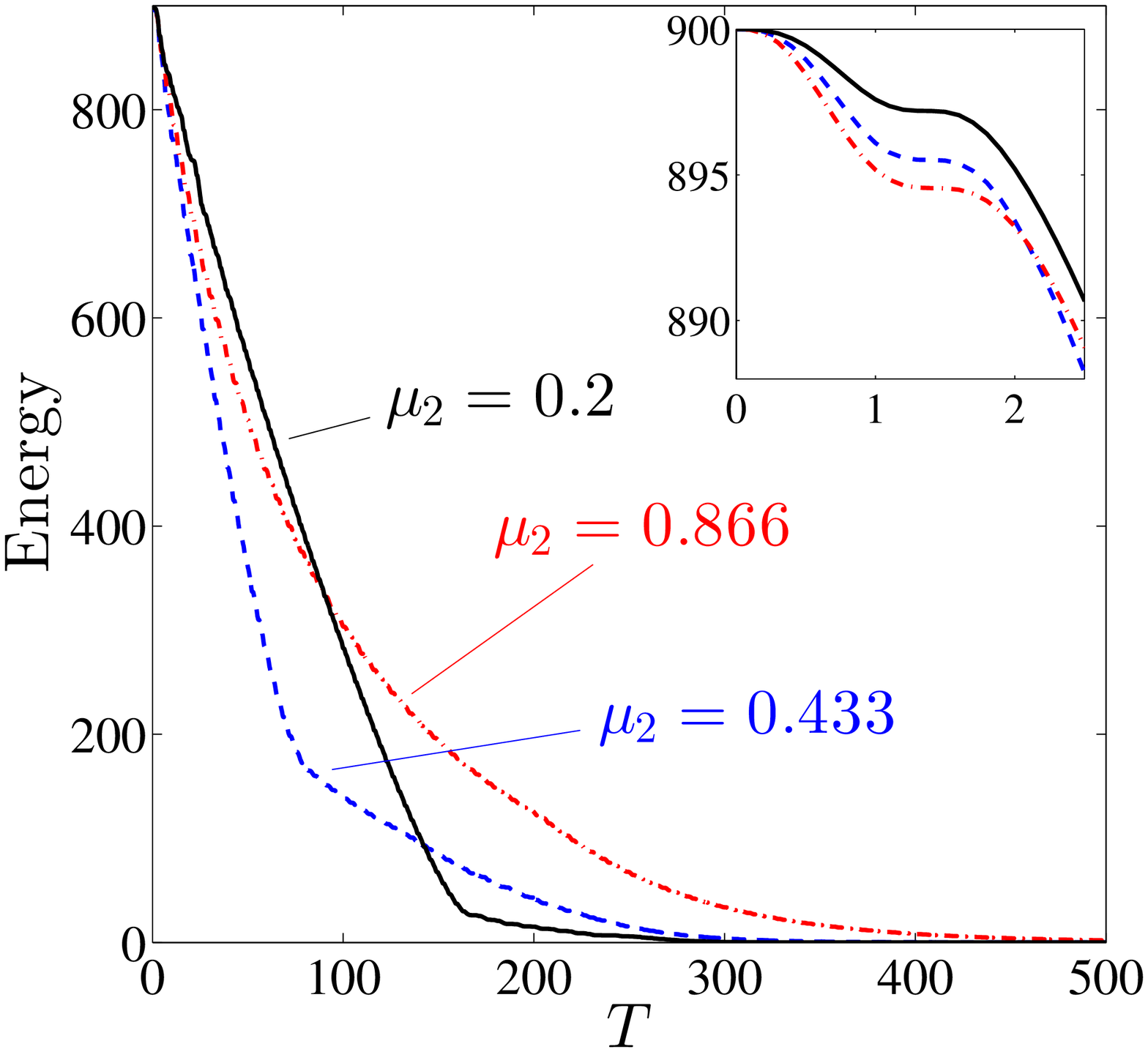}}
\put(0.02,0.33){(a)}
\put(0.46,0.33){(b)}
\par
\end{picture}
\end{centering}
\caption{\label{Power}(a) Estimated dissipation power based on the invariant manifold; (b) Energy decrement based on numerical simulations for the initial conditions $y_1=30$, $y_2=0$, $y_3=-\gamma/\sqrt{\lambda_3}$, $\dot y_1=\dot y_2=\dot y_3=0$. Parameter values $\gamma=0.707$, $\lambda_3=0.005$, $\varepsilon=0.05$, $\mu_2=0.2$, $0.433$ and $0.866$, as indicated in the figure.}
\end{figure}

The invariant manifold gives important information regarding the performance of the absorber. In fact the dissipation power of the absorber is equal to $P=2\mu_2\varepsilon \dot y_3^2\approx 2\mu_2 \varepsilon b_1^2$, i.e. it is proportional to $\mu_2 b_1^2$ (and to $\varepsilon$).
In this respect, Fig.~\ref{Power}(a) illustrates the quantity $\mu_2 b_1^2$, plotted according to Eq.~(\ref{many1}), for the three cases represented in Fig.~\ref{manifold}.
In the figure, going from right to left, i.e. for decreasing values of oscillation amplitude, initially the most damped absorber is performing the best (dash-dotted red curve, $\mu_2=0.866$).
For $a_1= 31.4$ the dashed blue curve ($\mu_2=0.433$) exceeds the red one, outperforming the other two absorbers. Finally, for $a_1=16.3$, the solid black curve ($\mu_2=0.2$) passes the dashed blue one, thus in this region the less damped absorber is the one with the best performance.

These analytical results are qualitatively confirmed by Fig.~\ref{Power}(b), where the energy decrement, computed from direct numerical simulations, is displayed.
The three curves refer to the time series shown in Fig.~\ref{manifold}.
In the inset of Fig.~\ref{Power}(b), it can be verified that, initially, the best performing absorber is the most damped one, since the red energy curve is the lowest one. For $T\approx 2$, the blue curve goes below the red one, meaning that the corresponding absorber outperforms the other ones. Finally, for $T\approx 142$, the solid black line, initially far above, goes below the other two energy curves, marking the overtaking of the less damped absorber with respect to the other two.
Qualitatively, the behavior of the system illustrated in Fig.~\ref{Power}(b), reflects the prediction given by Fig.~\ref{Power}(a).
Nevertheless, we note that a proper quantitative comparison cannot be performed between the two figures, since in Fig.~\ref{Power}(b) the energy decrement is represented with respect to time $T$, while Fig.~\ref{Power}(a) refers to the modal amplitude of the primary system.
This analysis suggests that an optimal damping coefficient does not exist in general, but its value should be chosen according to the most convenient dissipation strategy for the considered application.

\begin{figure}
\begin{centering}
\setlength{\unitlength}{\textwidth}
\begin{picture}(1,0.33)
\put(0.07,0.0){\includegraphics[trim = 10mm 10mm 10mm 10mm,width=0.4\textwidth]{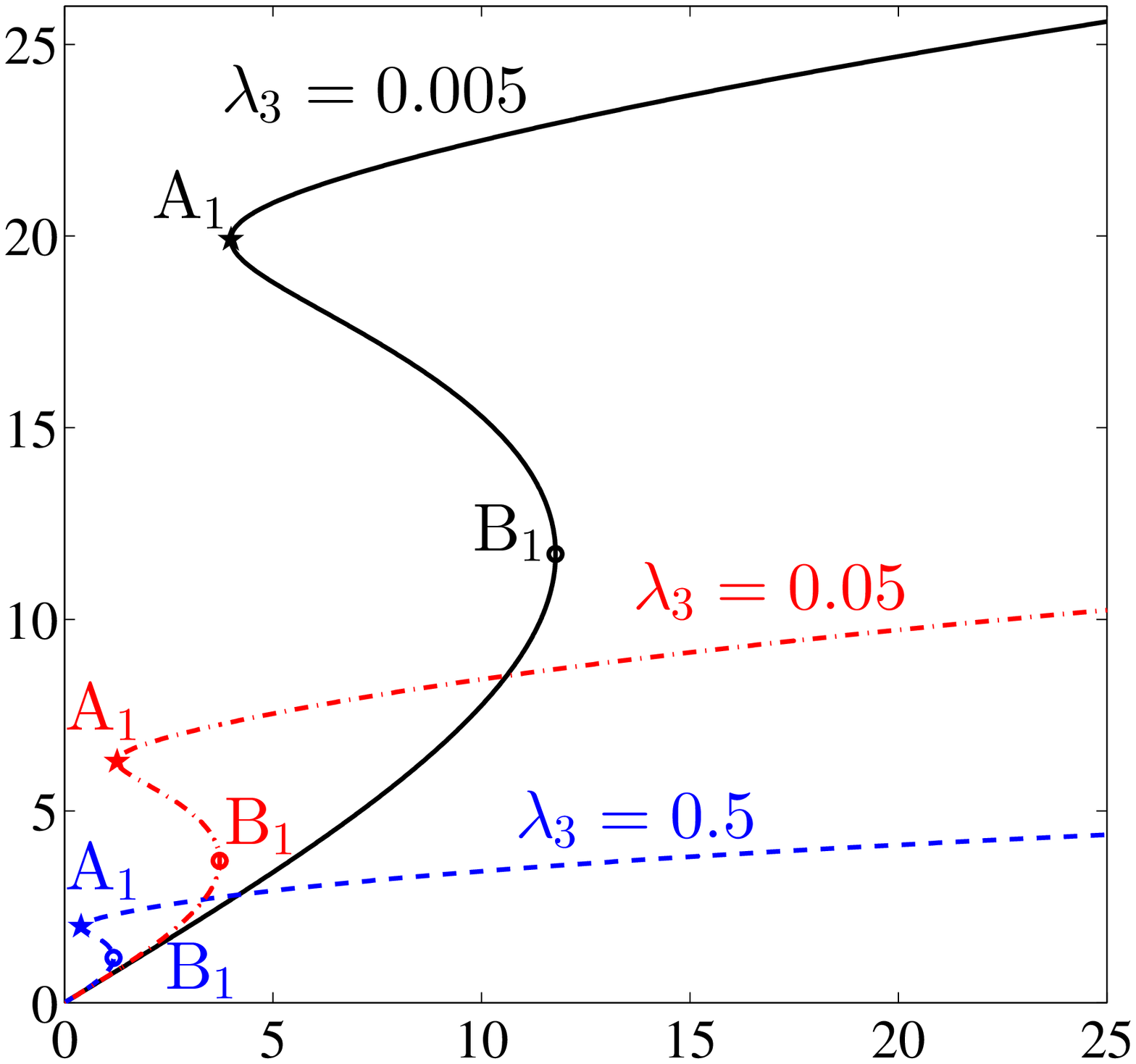}}
\put(0.51,0.0){\includegraphics[trim = 10mm 10mm 10mm 10mm,width=0.4\textwidth]{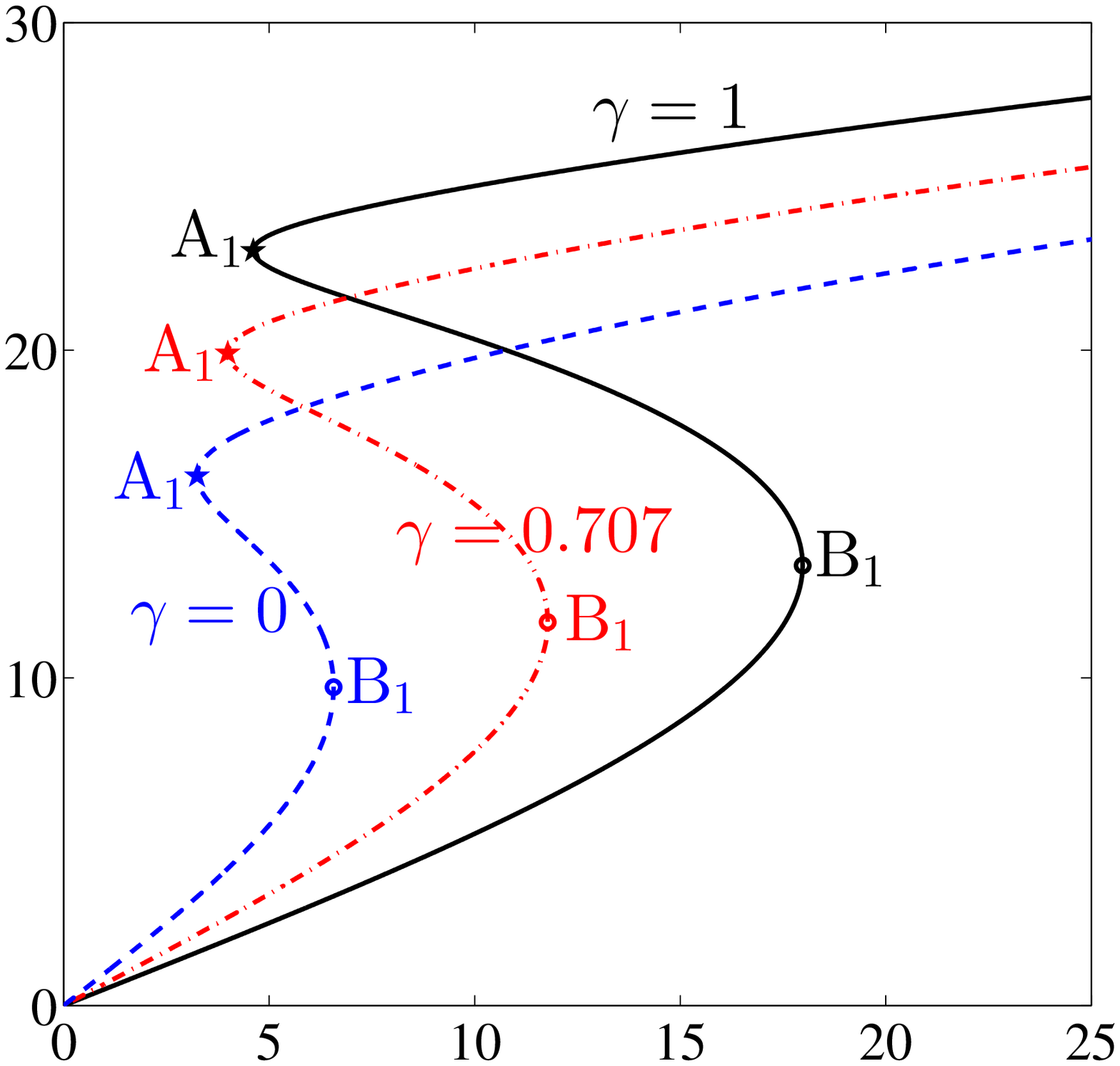}}
\put(0.05,0.33){(a)}
\put(0.49,0.33){(b)}
\put(0.06,0.19){$b_1$}
\put(0.5,0.19){$b_1$}
\put(0.275,-0.01){$a_1$}
\put(0.715,-0.01){$a_1$}
\par
\end{picture}
\end{centering}
\caption{\label{manifold_lambda}(a) Invariant manifolds for $\gamma=0.707$, $\mu_2=0.1$ and $\lambda_3=0.005$ (solid black line), $\lambda_3=0.05$ (dash-dotted red line) and $\lambda_3=0.5$ (dashed blue line). (b) Invariant manifolds for $\mu_2=0.1$, $\lambda_3=0.005$ and $\gamma=1$ (solid black line), $\gamma=0.707$ (dash-dotted red line) and $\gamma=0$ (dashed blue line).}
\end{figure}

The effect of variations of $\gamma$ and $\lambda_3$ on the manifold can be better understood analyzing how they influence coordinates of points A$_1$ and B$_1$.
As it can be clearly recognized from Eq.~(\ref{AB1}), both coordinates of A$_1$ (and of B$_1$) are inversely proportional to $\lambda_3$, thus variations of $\lambda_3$ will mainly scale the manifold in an inversely proportional way.
Increasing (decreasing) $\lambda_3$ the manifold is scaled down (up) as illustrated in Fig.~\ref{manifold_lambda}(a).
The results displayed in the figure suggest that low values of $\lambda_3$ can guarantee better performance of the absorber.
However, by decreasing $\lambda_3$, point A$_1$ is pulled to higher values of $a_1$, which means that the minimal energy threshold, typical for NES, is increased to higher energy values.
In fact, the $a_1$ coordinate of point A$_1$ marks the energy threshold of efficient behavior of the absorber (excluding the in-well and chaotic dynamics).

Also variations of $\gamma$ have important influence on the manifold (see Fig.~\ref{manifold_lambda}(b)).
Increases of $\gamma$ lift the upper branch of the manifold, resulting in better performance, however, the $a_1$ coordinate of A$_1$ increases as well.
At the same time, the $a_1$ coordinate of point B$_1$ grows significantly, which enlarges the bistable region.
If the system initial conditions fall within the bistable region, i.e. $a_1$ between $a_{1A}$ and $a_{1B}$, although the energy threshold is passed, the system might be attracted by the lower branch of the manifold, resulting in poor dissipation.

In order to understand more clearly the effect of $\gamma$ on the manifold, we approximate the value of the coordinates of A$_1$ and B$_1$ for large values of $\gamma$, i.e. \begin{equation}
\begin{split}
&\text A_1\approx\left(\frac{2\sqrt{2}\sqrt{1+12\mu_2^2-\sqrt{1-12\mu_2^2}}}{3\sqrt{3}}\frac{\gamma}{\sqrt{\lambda_3}},\frac{2\sqrt 3}{3}\frac{\gamma}{\sqrt{\lambda_3}}\right)\\
&\text B_1\approx\left(\frac{4}{9}\frac{\gamma^3}{\sqrt{\lambda_3}},\frac{2}{3}\frac{\gamma}{\sqrt{\lambda_3}}\right)
\end{split}\quad\text{ for }\gamma\gg 1. \label{ABapprox}
\end{equation}
As illustrated by Eq.~(\ref{ABapprox}), while A$_1$ grows linearly with $\gamma$ in both coordinates, B$_1$ grows linearly with $\gamma$ along $b_1$, but it increases proportionally to $\gamma^3$ along $a_1$. Thus, the bistable region is tremendously enlarged by big values of $\gamma$. In real applications this might compromise the robustness of the absorber.

Considering that $\lambda_3$ and $\gamma$ influence similarly the shape and amplitude of the manifold, it might be convenient to tune $\gamma$ in order to mitigate low amplitude dynamics, and optimize  $\lambda_3$ to mitigate the large amplitude dynamics, according to the requirements of the application in terms of target amplitude range and acceptable energy threshold.
As it will be shown later, by simultaneously acting on  $\gamma$ and $\lambda_3$, it is possible to significantly enlarge the amplitude range of efficiency of the absorber.
In passing, it should be noted that, if $\gamma$ is set to 0, the BNES is reduced to a purely cubic NES.

The same analysis can be analogously repeated considering the case when only the second mode is activated.
The resulting invariant manifold is described by \begin{equation}
a_2^2=\frac{b_2^2}{9}\left(\left(3+\gamma^2-\frac{3}{4}\lambda_3b_2^2\right)^2 +4\mu_2^2\right)
\end{equation}
where $a_1=b_1=0$.
For the sake of brevity, the analysis of the second mode invariant manifold is not performed here since qualitatively conclusions analogous to the previously discussed first mode case apply to the second one.

\subsection{Two modes dynamics}

We now consider the full manifold of Eqs.~(\ref{man1}) and (\ref{man2}), comprising initial conditions on both modes.
Rewriting Eqs.~(\ref{man1}) and (\ref{man2})  as \begin{equation}
\begin{split}
&\mathcal{A}_1=\mathcal{B}_1\cos\left(\beta_1-\beta_2\right)+\mathcal{C}_1\sin\left(\beta_1-\beta_2\right)\\
&\mathcal{A}_2=\mathcal{B}_2\cos\left(\beta_1-\beta_2\right)+\mathcal{C}_2\sin\left(\beta_1-\beta_2\right)
\end{split}
\end{equation}
it can be recognized that they define a system of equation, which is linear with respect to $\cos\left(\beta_1-\beta_2\right)$ and $\sin\left(\beta_1-\beta_2\right)$. Its solution is \begin{equation}
\cos\left(\beta_1-\beta_2\right)=\frac{\mathcal{A}_2 \mathcal{C}_1-\mathcal{A}_1 \mathcal{C}_2}{\mathcal{B}_2 \mathcal{C}_1-\mathcal{B}_1 \mathcal{C}_2}\quad\text{ and }\quad\sin\left(\beta_1-\beta_2\right)=\frac{\mathcal{A}_1 \mathcal{B}_2-\mathcal{A}_2 \mathcal{B}_1}{\mathcal{B}_2 \mathcal{C}_1-\mathcal{B}_1 \mathcal{C}_2},
\end{equation}
where $\mathcal{A}_{1,2}$, $\mathcal{B}_{1,2}$ and $\mathcal{C}_{1,2}$ are function of $a_1$, $a_2$, $b_1$, $b_2$, $\gamma$, $\mu_2$ and $\lambda_3$.

Adopting the Pythagorean identity $\cos\left(\beta_1-\beta_2\right)^2+\sin\left(\beta_1-\beta_2\right)^2=1$, the manifold is also described by the equation \begin{equation}
\left(\frac{\mathcal{A}_2 \mathcal{C}_1-\mathcal{A}_1 \mathcal{C}_2}{\mathcal{B}_2 \mathcal{C}_1-\mathcal{B}_1 \mathcal{C}_2}\right)^2+\left(\frac{\mathcal{A}_1 \mathcal{B}_2-\mathcal{A}_2 \mathcal{B}_1}{\mathcal{B}_2 \mathcal{C}_1-\mathcal{B}_1 \mathcal{C}_2}\right)^2-1=0 \label{cossin}
\end{equation}
which does not depend any more on the phase $\beta_1-\beta_2$ between the two modes.
Eq.~(\ref{cossin}) can be rewritten as \begin{equation}
\mathcal{S}_1a_1^4+\mathcal{S}_2a_2^4+\mathcal{S}_3a_1^2a_2^2+\mathcal{S}_4a_1^2+\mathcal{S}_5a_2^2+\mathcal{S}_6=0,\label{mani3D}
\end{equation}
where $\mathcal{S}_{1,...,6}$ are function of $b_1$, $b_2$, $\gamma$, $\mu_2$ and $\lambda_3$. Although they are too long to be explicitly written here, they can be easily computed with computer algebra.
Eq.~(\ref{mani3D}) defines a 3-dimensional hypersurface in the 4-dimensional space $a_1,a_2,b_1,b_2$.

\begin{figure}
\begin{centering}
\setlength{\unitlength}{\textwidth}
\begin{picture}(1,0.6)
\put(0.2,0.42){\includegraphics[trim = 10mm 10mm 10mm 10mm,width=0.5\textwidth]{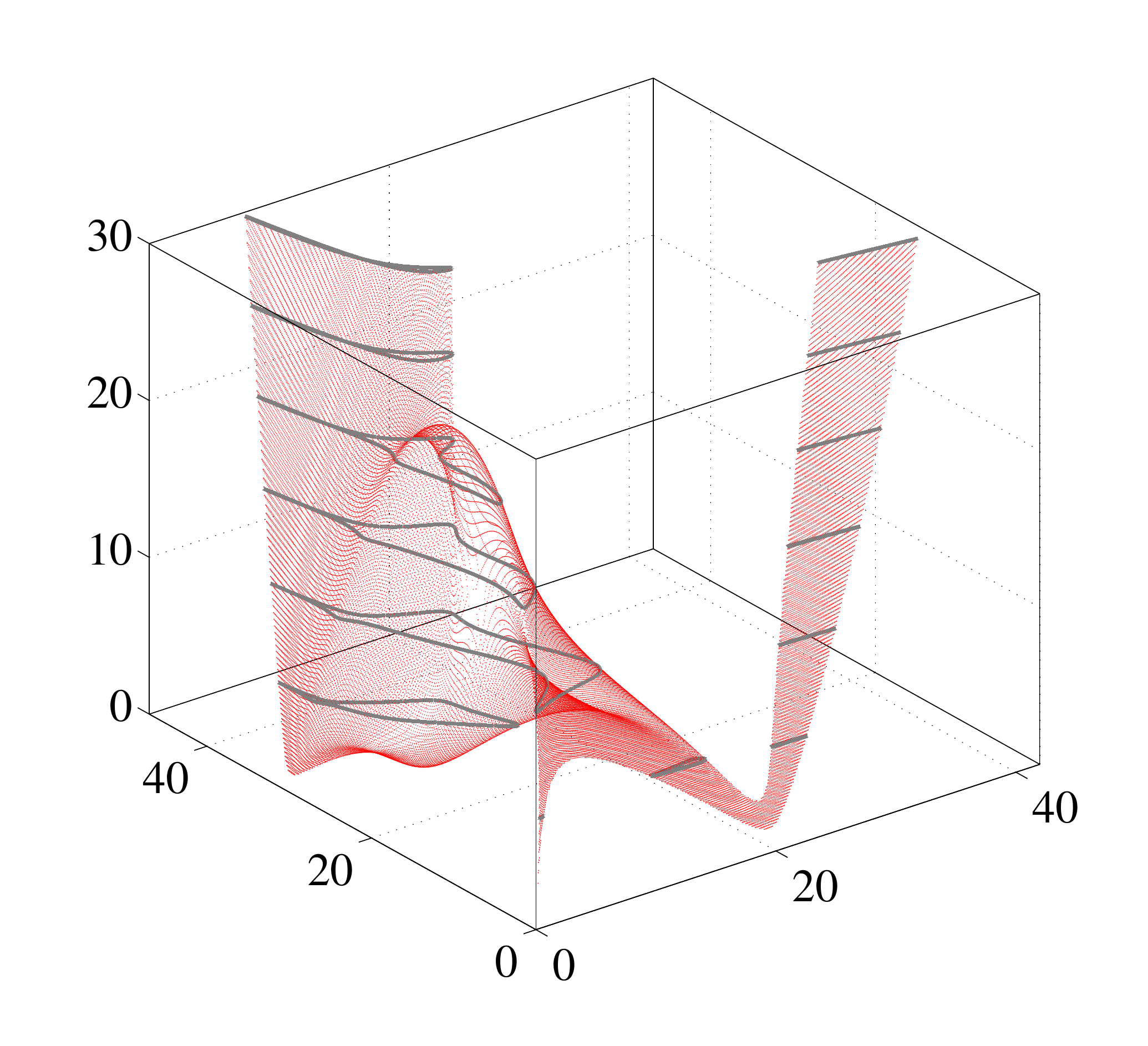}}
\put(0.06,0.02){\includegraphics[width=0.38\textwidth]{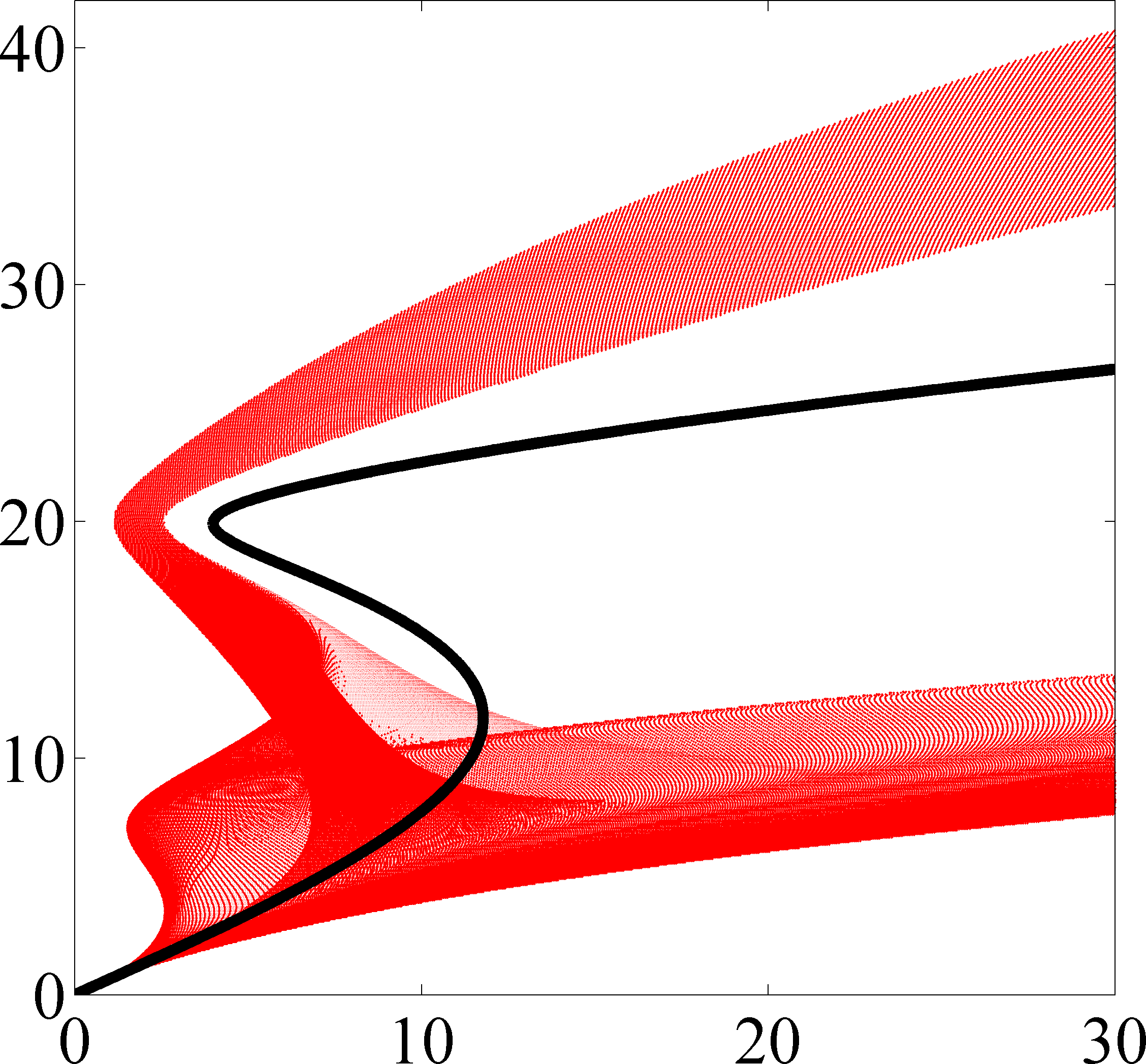}}
\put(0.51,0.02){\includegraphics[width=0.38\textwidth]{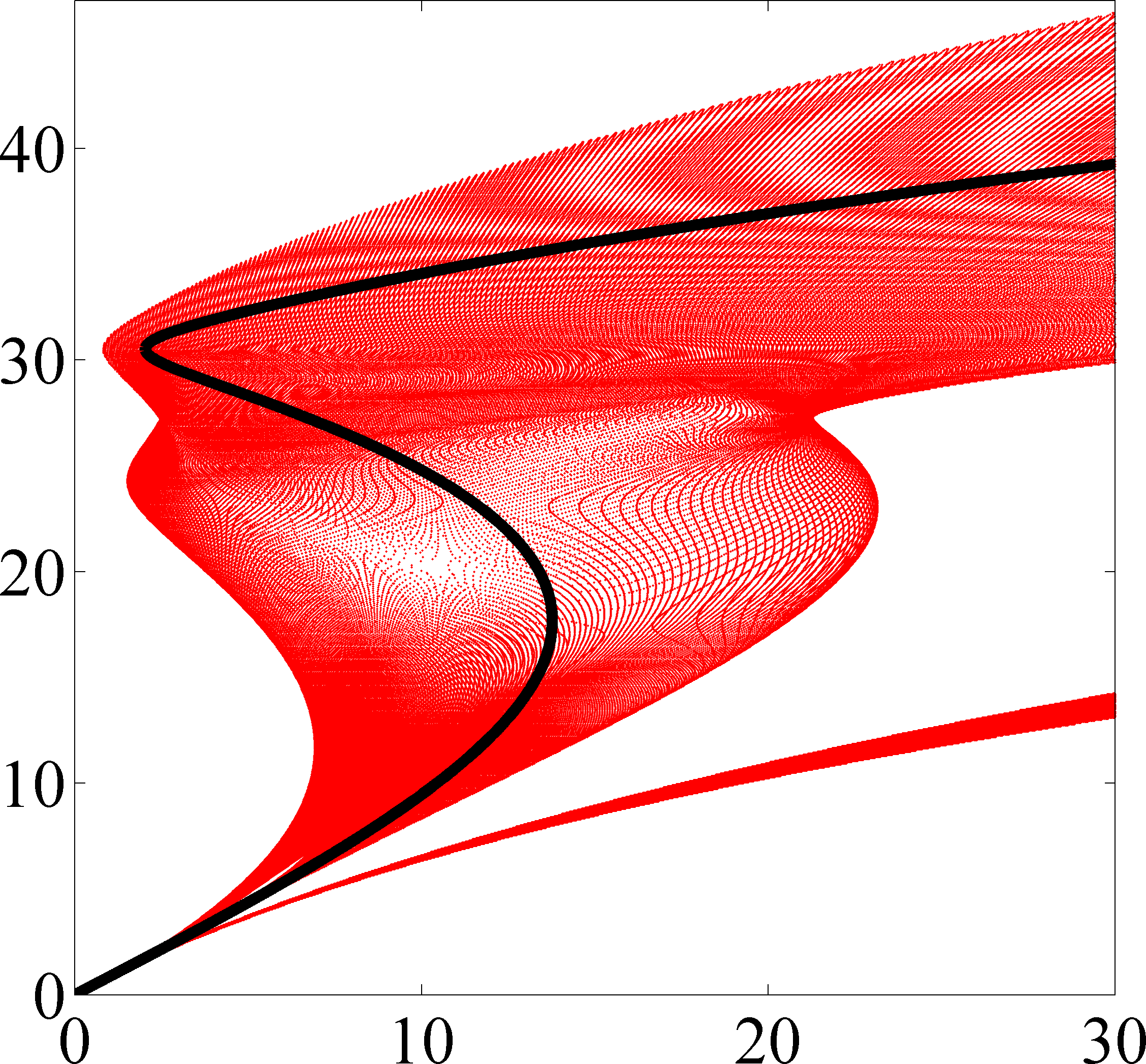}}
\put(0.18,0.78){(a)}
\put(0.02,0.36){(b)}
\put(0.48,0.36){(c)}
\put(0.16,0.67){$a_1,a_2$}
\put(0.27,0.47){$b_2$}
\put(0.6,0.45){$b_1$}
\put(0.02,0.21){$b_1$}
\put(0.48,0.21){$b_2$}
\put(0.23,0){$a_1,a_2$}
\put(0.68,0){$a_1,a_2$}
\par
\end{picture}
\end{centering}
\caption{\label{manifold3D}(a) Invariant manifold for $a_1=a_2$, $\gamma=0.707$, $\lambda_3=0.005$ and $\mu_2=0.1$; gray closed lines refer to constant values of $a_1$ and $a_2$. (b,c) Projection of the manifold on the $a_1,b_1$ and $a_2,b_2$ spaces; black solid lines: invariant manifolds for $a_2=0$ (b) and for $a_1=0$ (c).}
\end{figure}
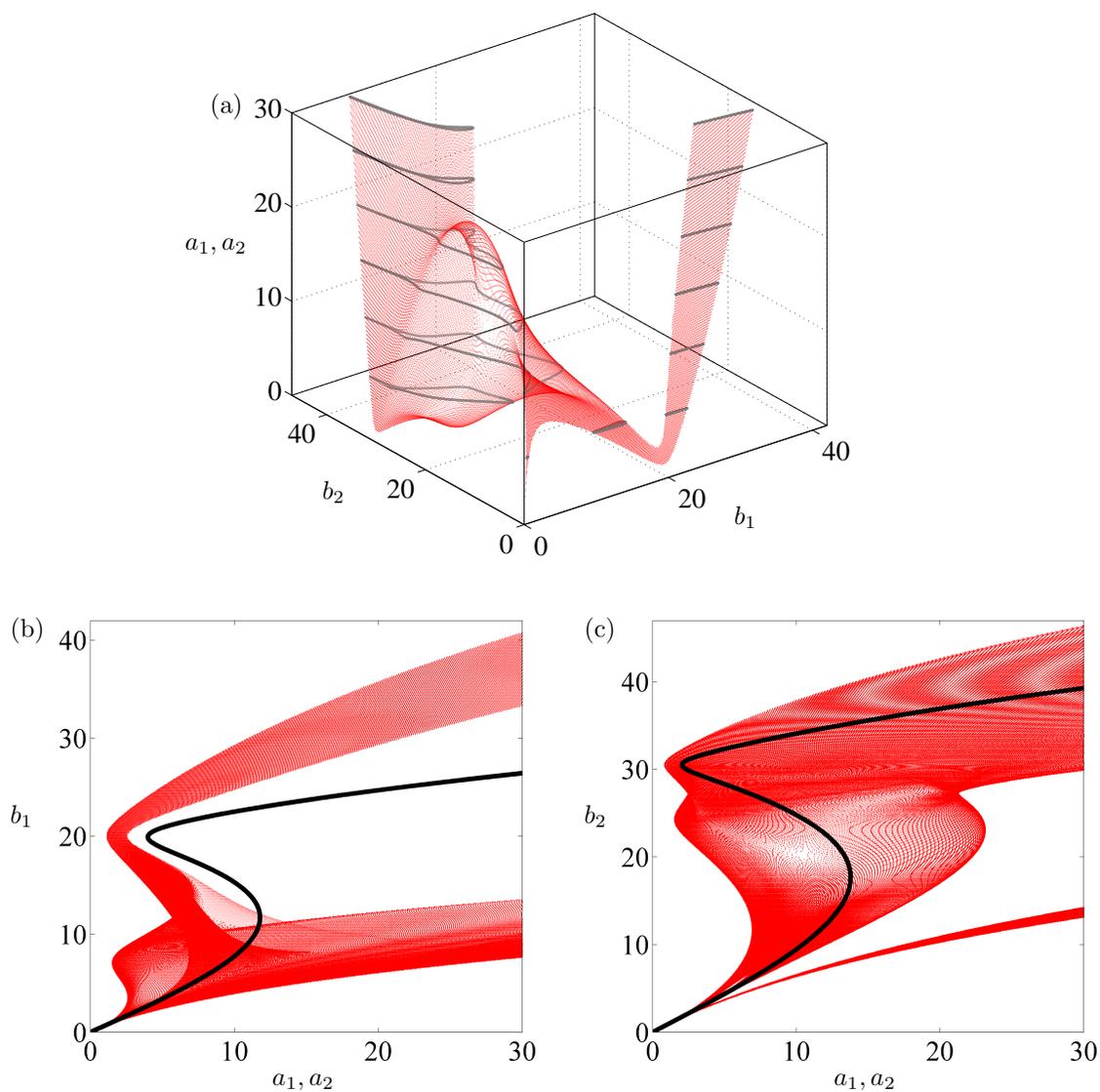

In order to plot the manifold, we impose a fixed ratio between $a_1$ and $a_2$, i.e. $a_2=pa_1$, which physically means that we fix the ratio between the modal amplitude of the two modes of the primary system.
The manifold is thus reduced to a 2-dimensional surface in the 3-dimensional space $a_1,b_1,b_2$ (or $a_2,b_1,b_2$).
The resultant manifold, for the parameter values $p=1$ ($a_1=a_2$), $\gamma=0.707$, $\lambda=0.005$ and $\mu_2=0.1$ is illustrated in Fig.~\ref{manifold3D}(a), while its projection on the $a_1,b_1$ and $a_2,b_2$ spaces are shown in Figs.~\ref{manifold3D}(b) and (c), respectively.
The grey closed lines in Fig.~\ref{manifold3D}(a) are sections for constant $a_1$ and $a_2$ values.
The black solid lines in Figs.~\ref{manifold3D}(b) and (c) depict the invariant manifold for the single mode case, for the first and the second mode, respectively.

Although the two modes manifold has a quite involved geometry, which prevents us from using it for design purposes, it can still give some insight about the system dynamics.
Analyzing its projection on the $a_1,b_1$ and $a_2,b_2$ spaces, it can be observed that in both cases it resembles the shape of the single mode manifolds (black lines), although it presents two separated bundles.
The upper bundle in Fig.~\ref{manifold3D}(b) corresponds to the lower bundle in Fig.~\ref{manifold3D}(c) and vice-versa.
The upper bundle of Fig.~\ref{manifold3D}(b), engaging higher values of $b_1$, entails a better dissipation of the energy associated to the first mode; contrarily, the lower bundle of Fig.~\ref{manifold3D}(b) involves an opposite scenario.

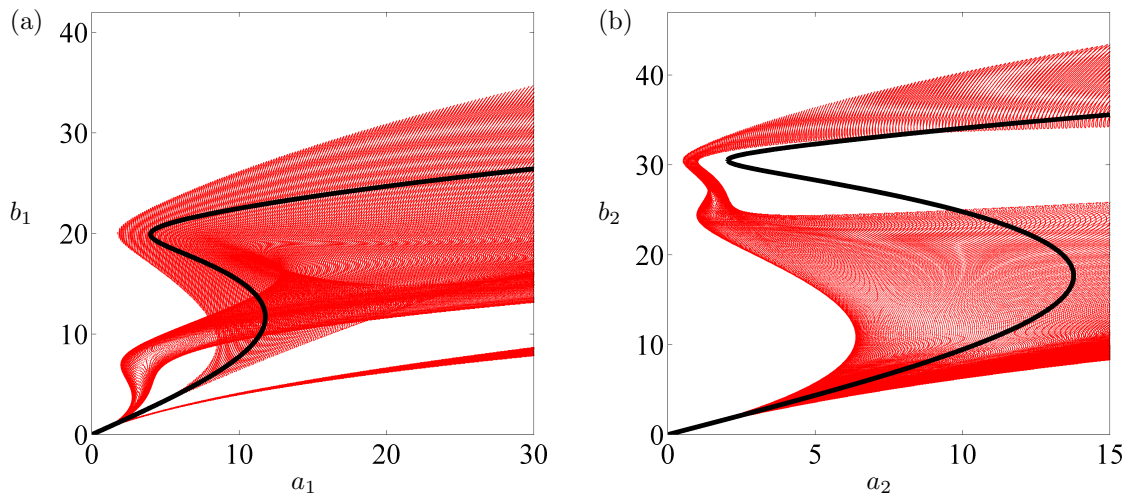
\begin{figure}
\begin{centering}
\setlength{\unitlength}{\textwidth}
\begin{picture}(1,0.3)
\put(0.09,0.02){\includegraphics[width=0.38\textwidth]{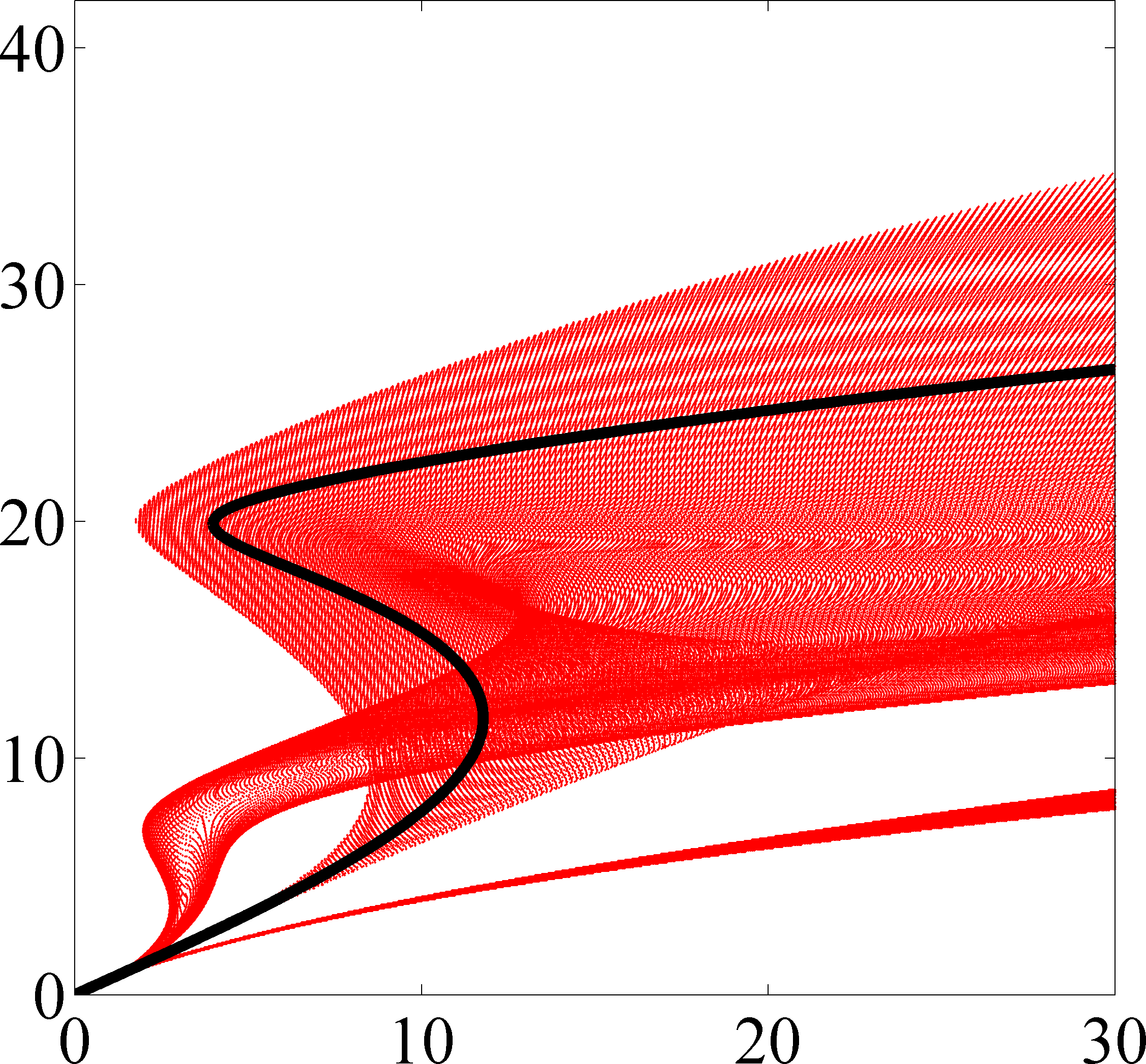}}
\put(0.54,0.02){\includegraphics[width=0.38\textwidth]{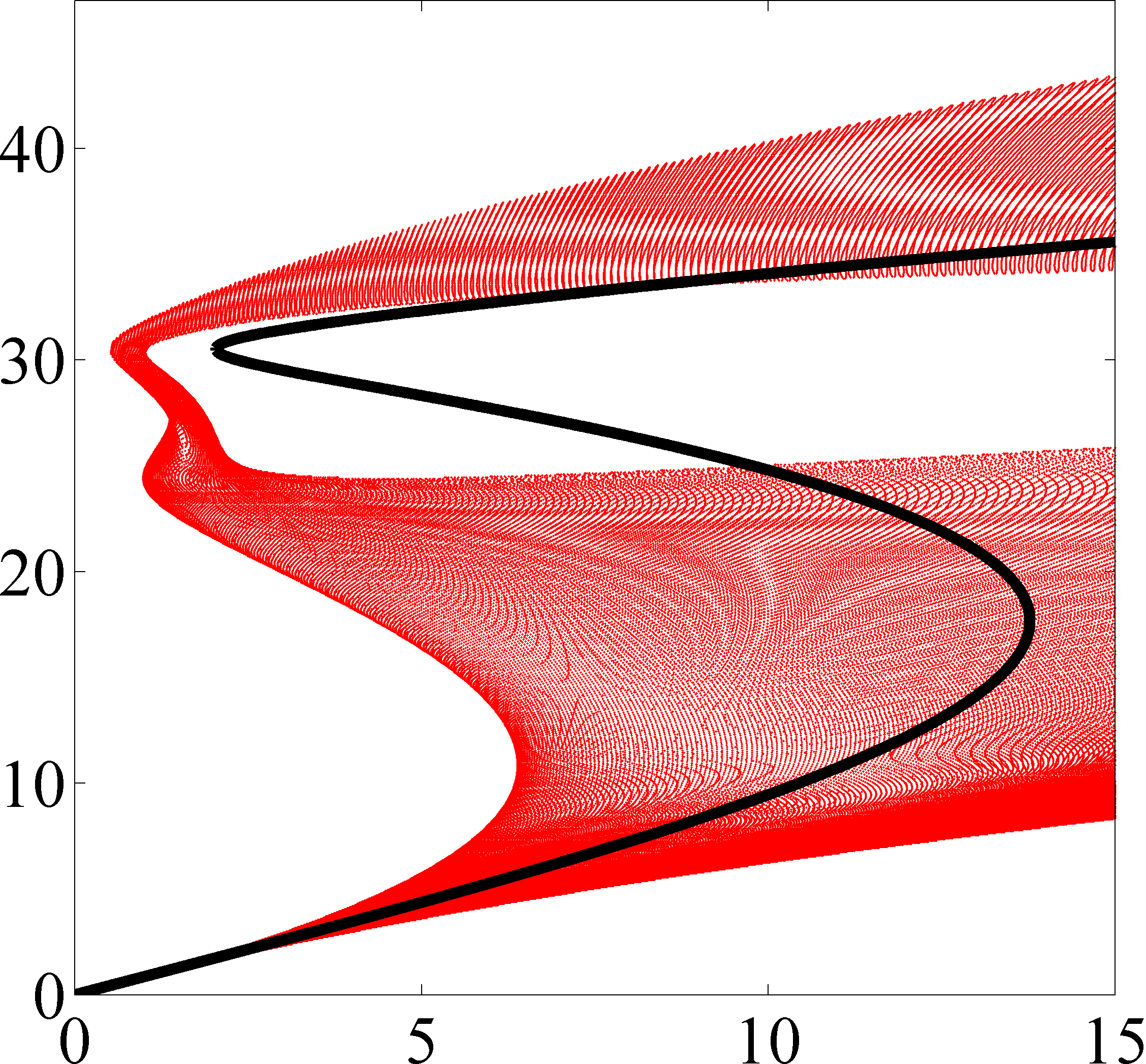}}
\put(0.05,0.36){(a)}
\put(0.51,0.36){(b)}
\put(0.05,0.21){$b_1$}
\put(0.51,0.21){$b_2$}
\put(0.27,0){$a_1$}
\put(0.72,0){$a_2$}
\par
\end{picture}
\end{centering}
\caption{\label{manifold_p05}Projection of the two modes manifold on the $a_1,b_1$ (a) and $a_2,b_2$ (b) spaces; black solid lines: invariant manifolds for $a_2=0$ (a) and for $a_1=0$ (b). Parameter values $p=0.5$ ($a_2=0.5a_1$), $\gamma=0.707$, $\lambda_3=0.005$ and $\mu_2=0.1$.}
\end{figure}

Figs.~\ref{manifold_p05}(a) and (b) illustrate the projections of the manifold, now obtained for a different ratio between $a_2$ and $a_1$ ($p=0.5$).
Comparing Figs.~\ref{manifold3D}(b) and (c) with Figs.~\ref{manifold_p05}(a) and (b), it can be recognized that, although the shape of the manifold is similar, the two bundles present very different width.
Assuming that the width of a bundle is directly related to the probability that the system dynamics converges towards it, it might be deduced that the manifold illustrated in Fig.~\ref{manifold3D} involves a better dissipation of the energy on the second mode than on the first mode.
However, in this case, the ratio between $a_2$ and $a_1$ would reduce during the transient dynamics, leading to a scenario similar to the one illustrated in Fig.~\ref{manifold_p05}, which instead involves a better dissipation of the energy on the first mode (the upper bundle in Fig.~\ref{manifold_p05}(a) is much more width than the lower one).
Therefore it can be inferred that this mechanism facilitates a balanced dissipation of the energy on the two modes.

\section{Numerical validation of the absorber performance}\label{numerical_validation}

\begin{figure}
\begin{centering}
\setlength{\unitlength}{\textwidth}
\begin{picture}(1,1.2)
\put(-0.07,1.2){\includegraphics[trim = 0mm 210mm 5mm 5mm,width=\textwidth]{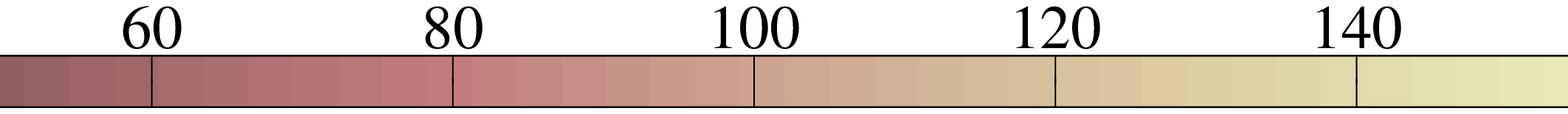}}
\put(0.04,0.82){\includegraphics[trim = 10mm 10mm 10mm 10mm,width=0.4\textwidth]{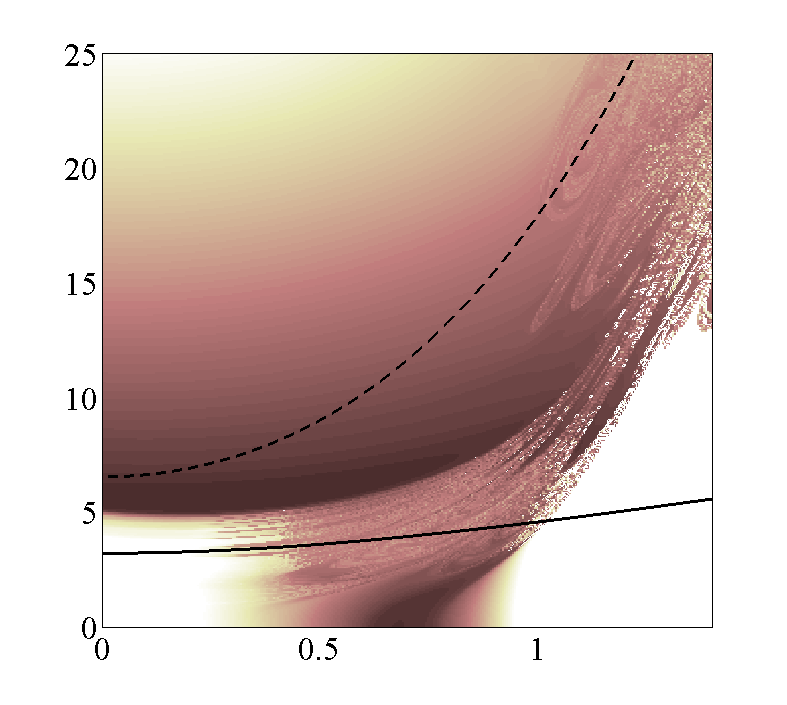}}
\put(0.46,0.82){\includegraphics[trim = 10mm 10mm 10mm 10mm,width=0.4\textwidth]{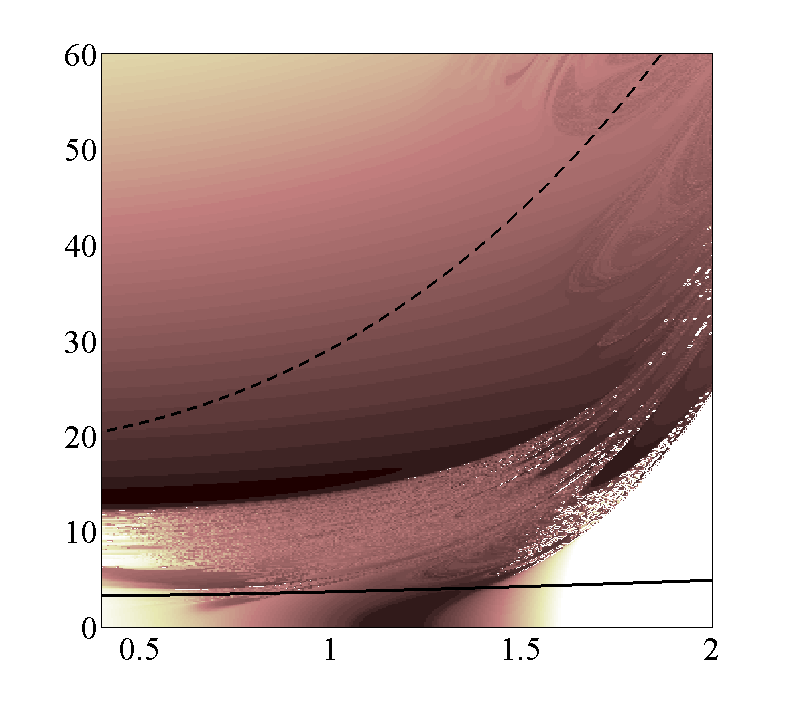}}
\put(0.04,0.42){\includegraphics[trim = 10mm 10mm 10mm 10mm,width=0.4\textwidth]{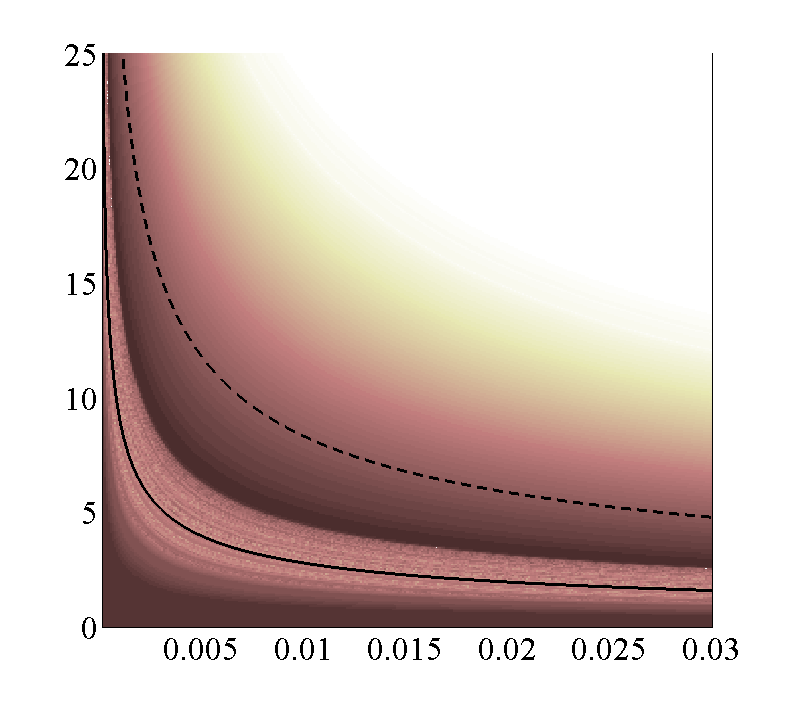}}
\put(0.46,0.42){\includegraphics[trim = 10mm 10mm 10mm 10mm,width=0.4\textwidth]{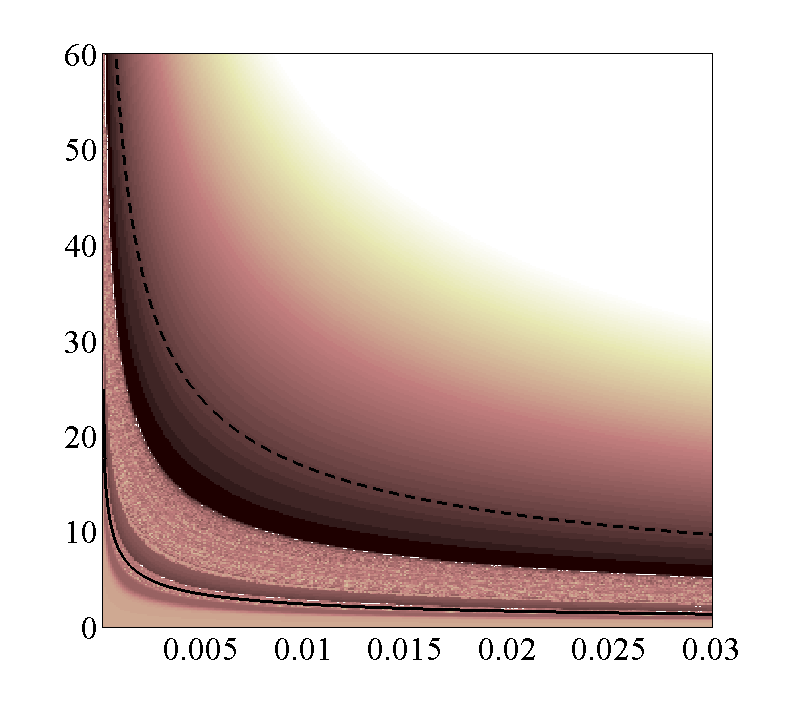}}
\put(0.04,0.02){\includegraphics[trim = 10mm 10mm 10mm 10mm,width=0.4\textwidth]{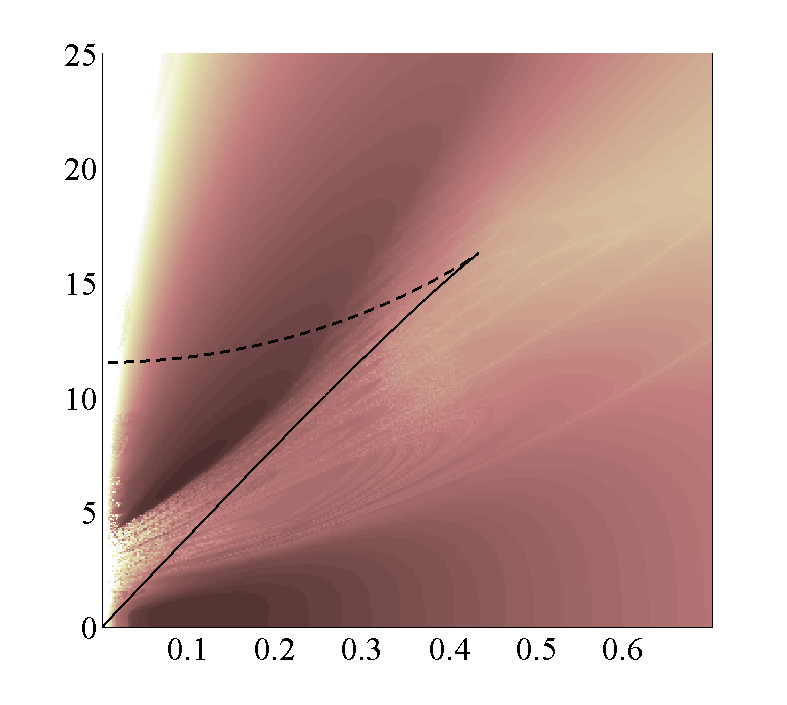}}
\put(0.46,0.02){\includegraphics[trim = 10mm 10mm 10mm 10mm,width=0.4\textwidth]{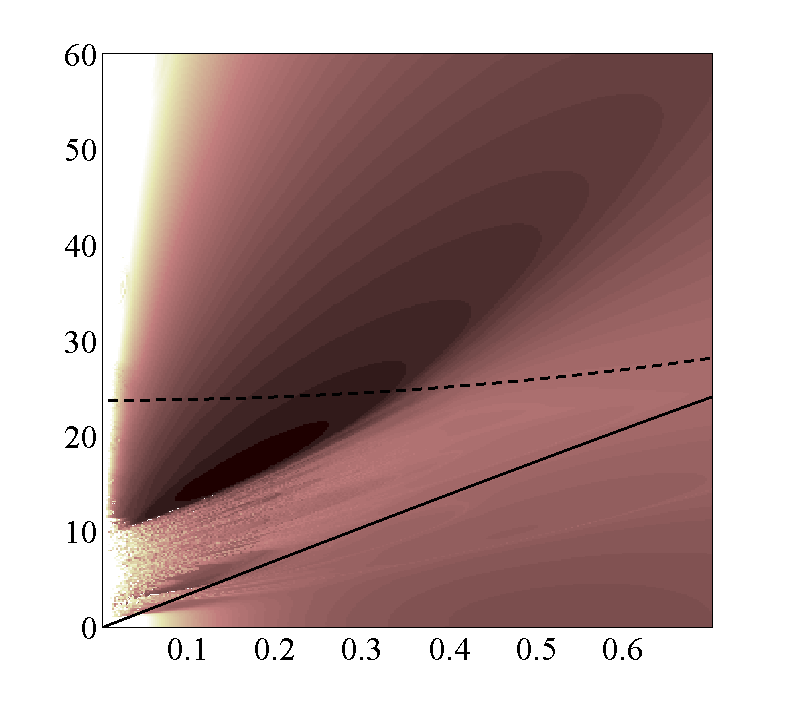}}
\put(0.37,0.915){A$_1$}
\put(0.3,1.1){B$_1$}
\put(0.79,0.875){A$_2$}
\put(0.72,1.1){B$_2$}
\put(0.37,0.47){A$_1$}
\put(0.37,0.515){B$_1$}
\put(0.79,0.455){A$_2$}
\put(0.79,0.505){B$_2$}
\put(0.235,0.19){A$_1$}
\put(0.235,0.24){B$_1$}
\put(0.79,0.135){A$_2$}
\put(0.79,0.2){B$_2$}
\put(0.02,1.21){$T$}
\put(0.02,1.16){(a)}
\put(0.44,1.16){(b)}
\put(0.02,0.76){(c)}
\put(0.44,0.76){(d)}
\put(0.02,0.36){(e)}
\put(0.44,0.36){(f)}
\put(0.02,1.01){$v_0$}
\put(0.44,1.01){$v_0$}
\put(0.02,0.61){$v_0$}
\put(0.44,0.61){$v_0$}
\put(0.02,0.21){$v_0$}
\put(0.44,0.21){$v_0$}
\put(0.25,0.8){$\gamma$}
\put(0.67,0.8){$\gamma$}
\put(0.25,0.4){$\lambda_3$}
\put(0.67,0.4){$\lambda_3$}
\put(0.25,0){$\mu_2$}
\put(0.67,0){$\mu_2$}
\par
\end{picture}
\end{centering}
\caption{\label{timemap}Dissipation time (70~\% of initial energy) for initial conditions $y_1=y_2=0$, $y_3=-\gamma/\sqrt{\lambda_3}$, $\dot y_1=v_0$, $\dot y_2=\dot y_3=0$ (first mode) (a,c,e) and $y_1=y_2=0$, $y_3=-\gamma/\sqrt{\lambda_3}$, $\dot y_2=v_0$, $\dot y_1=\dot y_3=0$ (second mode) (b,d,f). (a,b) $\lambda_3=0.005$, $\mu_2=0.1$; (c,d) $\gamma=0.707$, $\mu_2=0.1$; (e,f) $\lambda_3=0.005$, $\gamma=0.707$. $\varepsilon=0.05$ in all plots. Solid lines indicate points A$_1$ and A$_2$ of the single mode manifolds; dashed lines indicate points B$_1$ and B$_2$. Simulations were limited to 200 time intervals.}
\end{figure}

Extensive numerical simulations of the system subject to impulsive excitation are performed in this section.
The results are summarized in Fig.~\ref{timemap}, where the color maps indicate the energy dissipation time.
In the figure, points A$_1$ and B$_1$ of the first mode manifold, and points A$_2$ and B$_2$, analogous of the second mode manifold, are also depicted, in order to interpret numerical results with respect to analytical insights.

Figures~\ref{timemap}(a) and (b) illustrate the dissipation time of 70~\% of the initial energy, for various values of $\gamma$ and of initial energy.
Fig.~\ref{timemap}(a) refers to an impulsive excitation of the first mode, while Fig.~\ref{timemap}(b) to an impulsive excitation of the second mode.
For low amplitudes, both figures clearly show the importance of a proper tuning of $\gamma$, such that the absorber is in 1:1 resonance with the primary system. The calculated optimal values of $\gamma$ are $\gamma=1/\sqrt{2}=0.707$, for the first mode, and $\gamma=\sqrt{3/2}=1.225$ for the second.
It is therefore necessary to choose one of the modes to be targeted.
By doing so, we inherit the TMD unavoidable narrow band limitation. However, we restrict such limitation merely to the low amplitude in-well dynamics.
For increasing impulsive energy ($v_0\approx2$ for Fig.~\ref{timemap}(a)), the dark blue region of rapid energy decrement moves slightly to the right.
This is due to the softening effect occurring in-well, thus, to meet an efficient 1:1 resonance, the optimal value of $\gamma$ slightly increases.
For larger values of initial impulse, the system enters the chaotic region.
Dissipation times increase, but the optimal values of $\gamma$ are still around $1/\sqrt 2$ and $\sqrt{3/2}$ for the first and the second mode, respectively. However, the range of effective $\gamma$ values enlarges.

For even larger energy content, the contribution of the double-well becomes negligible and the dynamics is controlled by the shape of the manifold.
In the region between the solid (marking points A$_{1,2}$) and the dashed lines (marking points B$_{1,2}$), according to the invariant manifold, the absorber might be attracted either to the lower or to the higher branch of the manifold, as in Fig.~\ref{manifold}(a).
Intuitively we expect that, the closer we are to the dashed line, the more robust is the large amplitude attractor.
This is verified by numerical results, which illustrate that there is a sort of border, approximately in between the solid and the dashed lines, which marks the separation between fast and slow dissipation time of the absorber.
This result confirms once more the relevance of the invariant manifold for design purposes.
For very large initial energy, the value of $\gamma$ becomes less and less relevant with respect to dissipation time.
Both Figs.~\ref{timemap}(a) and (b) have the same qualitative behavior.

Figures~\ref{timemap}(c) and (d) depicts the energy dissipation time, for various values of $\lambda_3$, for excitation on the first and on the second mode, respectively.
$\gamma$ is tuned according to the first mode in both figures, i.e. $\gamma=1/\sqrt 2$. Thus, at low amplitude, where the effect of $\lambda_3$ is negligible, the dissipation is faster for the first mode than for the second.
By investigating higher energy levels, it can be immediately recognized how solid and dashed lines follow the numerical trend of the absorber behavior, as already noted for Figs.~\ref{timemap}(a) and (b).
As pointed out in Section 3.1, A$_{1,2}$ and B$_{1,2}$ have coordinates inversely proportional to the square root of $\lambda_3$, thus the absorber works efficiently at high energy level for low values of $\lambda_3$. On the contrary, for large values of $\lambda_3$, the absorber becomes rapidly inefficient.
The value of $\lambda_3$ should thus be chosen according to the target energy level for both modes.

Finally, Figs.~\ref{timemap}(e) and (f) refer to variations of damping $\mu_2$.
For $\mu_2$ that tends to 0, regardless of initial energy, dissipation time grows unboundedly, since the system loses its only dissipation term.
At low amplitude, first mode motions are effectively dissipated for a large range of damping values, on the contrary, to dissipate second mode motions, a relatively high damping is required, which enlarges the dissipation bandwidth.
Considering higher initial energy, the chaotic regime is recognizable, marked by scattered colors in the figure, which is smoothed for high values of $\mu_2$.
Although the region of rapid dissipation through large periodic motions is clearly identifiable for both modes (starting at $v_0\approx5$ for Fig.~\ref{timemap}(e) and at $v_0\approx10$ for Fig.~\ref{timemap}(f)), the relation between numerical results and the solid and dashed lines is not immediately recognizable.
However, this area tends to widen and to move to higher values of $v_0$ for increasing values of $\mu_2$. This behavior was already observed in Fig.~\ref{Power}(a), where the region of rapid dissipation corresponds to a bump of the dissipation power curve.
Although an accurate tuning of $\mu_2$ is advisable for a practical realization of a TBNES, as a general rule, we notice that high values of damping allow to increase the amplitude and frequency bandwidth of the absorber.

\section{Comparison among TBNES, TMD and NES}\label{Sect_comparison}

The analytical study performed for the TBNES, can be readily extended to the TMD and the NES (although the TMD, being linear, can be more efficiently studied in different ways).
As already mentioned, the NES can be considered as a BNES in which $\gamma=0$.
The equation of motion of the same primary system under study, with an attached TMD, are obtained by changing the sign of $\gamma^2$ and imposing $\lambda_3=0$.

\begin{figure}
\begin{centering}
\setlength{\unitlength}{\textwidth}
\begin{picture}(1,0.33)
\put(0.07,0.0){\includegraphics[trim = 10mm 10mm 10mm 10mm,width=0.4\textwidth]{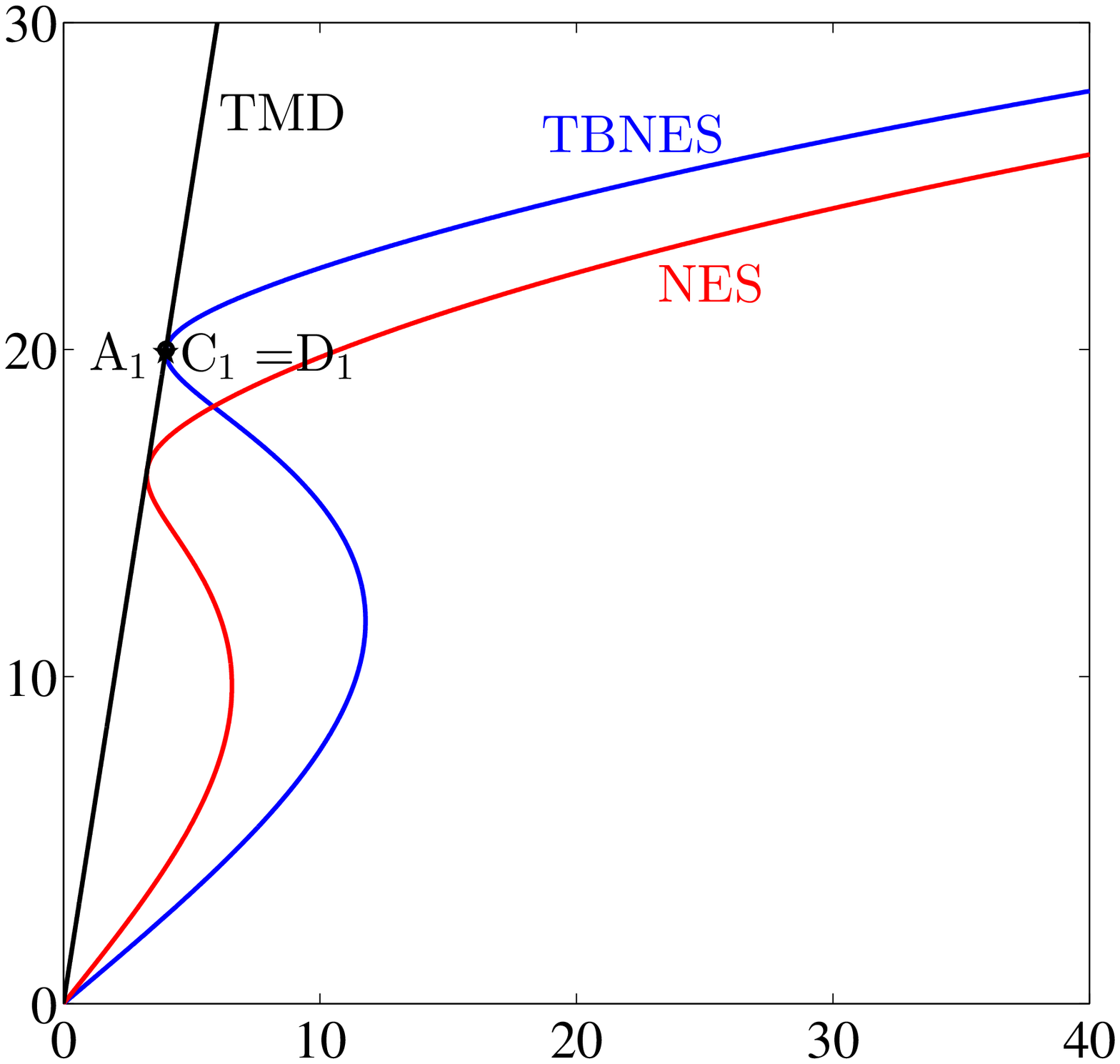}}
\put(0.51,0.0){\includegraphics[trim = 10mm 10mm 10mm 10mm,width=0.4\textwidth]{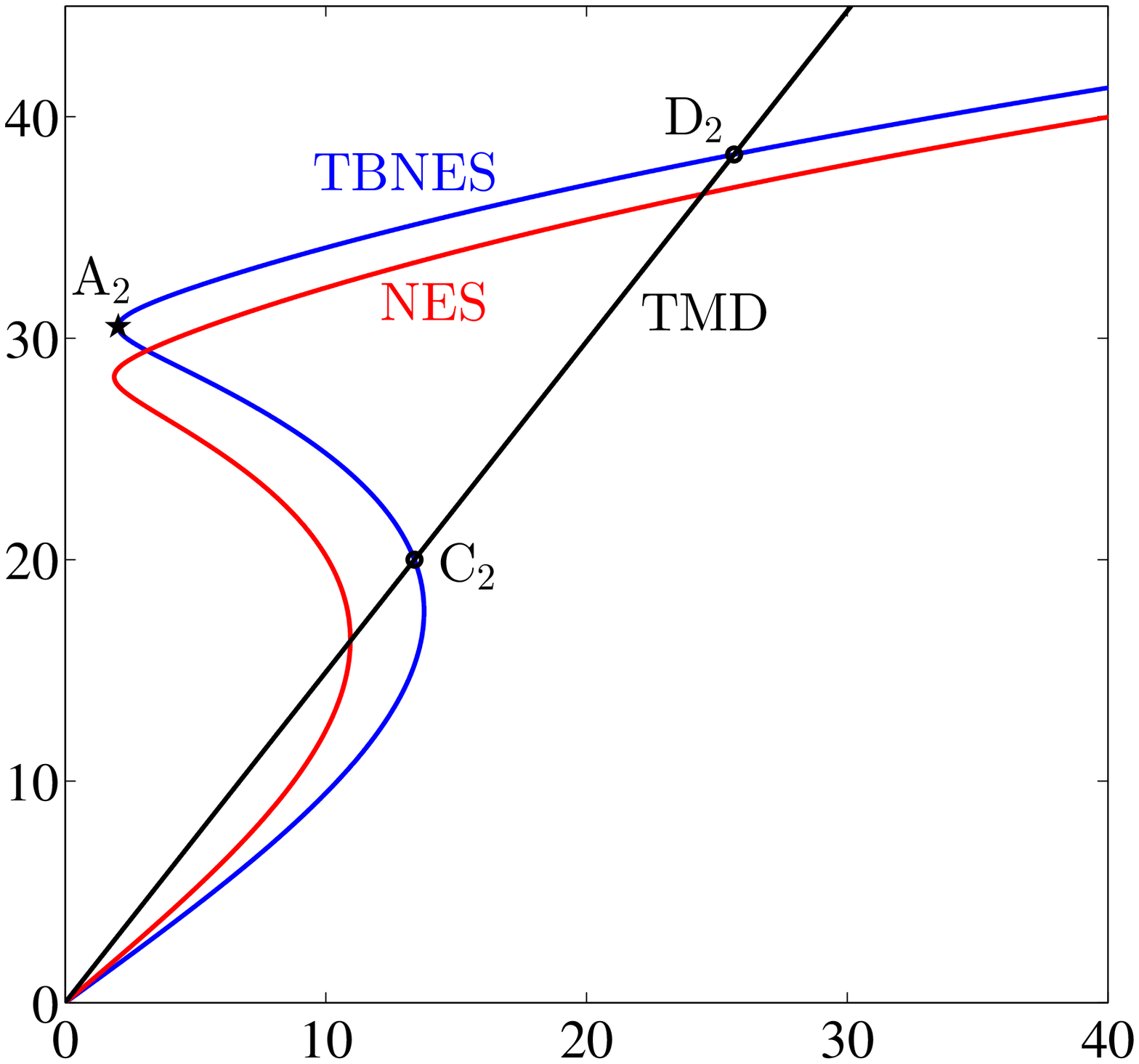}}
\put(0.05,0.33){(a)}
\put(0.49,0.33){(b)}
\put(0.06,0.19){$b_1$}
\put(0.5,0.19){$b_2$}
\put(0.275,-0.01){$a_1$}
\put(0.715,-0.01){$a_2$}
\par
\end{picture}
\end{centering}
\caption{\label{manifold_vari_abs}Invariant manifolds for different absorbers and the parameter values $\gamma=0.707$, $\gamma_{TMD}=1$, $\mu_2=0.1$ and $\lambda_3=0.005$. (a) first mode, (b) second mode.}
\end{figure}

The equations describing the invariant manifolds, for the first and second mode, of the TBNES, TMD and NES are \begin{eqnarray}
\text{TBNES}_1:&&a_1^2=b_1^2\left(\left(1+\gamma^2-\frac{3}{4}\lambda_3b_1^2\right)^2 +4\mu_2^2\right)\\
\text{TMD}_1:&&a_1^2=b_1^2\left(\left(1-\gamma_{TMD}^2\right)^2 +4\mu_2^2\right)\\
\text{NES}_1:&&a_1^2=b_1^2\left(\left(1-\frac{3}{4}\lambda_3b_1^2\right)^2 +4\mu_2^2\right)\\
\text{TBNES}_2:&&a_2^2=\frac{b_2^2}{9}\left(\left(3+\gamma^2-\frac{3}{4}\lambda_3b_2^2\right)^2 +4\mu_2^2\right)\\
\text{TMD}_2:&&a_2^2=\frac{b_2^2}{9}\left(\left(3-\gamma_{TMD}^2\right)^2 +4\mu_2^2\right)\\
\text{NES}_2:&&a_2^2=\frac{b_2^2}{9}\left(\left(3-\frac{3}{4}\lambda_3b_2^2\right)^2 +4\mu_2^2\right).
\end{eqnarray}
It can be noticed that for the TMD the manifolds are reduced to a straight line in the $a_1,b_1$ and $a_2,b_2$ spaces.
Furthermore, even its complete manifold involving both modes is decoupled.
The different manifolds are plotted in Fig.~\ref{manifold_vari_abs} for the parameter values $\gamma=0.707$, $\gamma_{TMD}=1$, $\mu_2=0.1$ and $\lambda_3=0.005$.

In order to compare the performance of the TBNES with respect to the TMD, we show in the same plot the relevant manifolds.
Points C$_{1,2}$ and D$_{1,2}$ in Fig.~\ref{manifold_vari_abs} mark the intersection between the TMD and the TBNES manifolds. Their coordinates, in the $a_1,b_1$ and $a_2,b_2$ space, can be easily obtained analytically and are given by
\begin{equation}
\begin{split}
\text C_1&=\left(2\sqrt{\frac{\left(\gamma^2+\gamma_{TMD}^2\right)\left(\left(1-\gamma_{TMD}^2\right)^2+4\mu_2^2\right)}{3\lambda_3}},2\sqrt{\frac{\gamma^2+\gamma_{TMD}^2}{3\lambda_3}}\right)\\
\text D_1&=\left(2\sqrt{\frac{\left(2+\gamma^2-\gamma_{TMD}^2\right)\left(\left(1-\gamma_{TMD}^2\right)^2+4\mu_2^2\right)}{3\lambda_3}},2\sqrt{\frac{2+\gamma^2-\gamma_{TMD}^2}{3\lambda_3}}\right)\\
\text C_2&=\left(\frac{2}{3}\sqrt{\frac{\left(\gamma^2+\gamma_{TMD}^2\right)\left(\left(3-\gamma_{TMD}^2\right)^2+4\mu_2^2\right)}{3\lambda_3}},2\sqrt{\frac{\gamma^2+\gamma_{TMD}^2}{3\lambda_3}}\right)\\
\text D_2&=\left(\frac{2}{3}\sqrt{\frac{\left(6+\gamma^2-\gamma_{TMD}^2\right)\left(\left(3-\gamma_{TMD}^2\right)^2+4\mu_2^2\right)}{3\lambda_3}},2\sqrt{\frac{6+\gamma^2-\gamma_{TMD}^2}{3\lambda_3}}\right).
\end{split}
\end{equation}

The TBNES is expected to provide better performance than the TMD when its manifold branch is higher than the TMD one, which occurs between point A$_{1,2}$ and D$_{1,2}$.
We consider a TMD and a TBNES tuned according to the first mode, thus $\gamma_{TMD}=1$ and $\gamma=0.707$.
Regarding the first mode, the TMD manifold is always higher than the TBNES one.
It can be verified that in this condition (for $\gamma_{TMD}=1$), for any parameter set of the TBNES, the TMD manifold is tangent to the TBNES manifold, with the former, as expected, always laying above the latter.
This result confirms the well known excellent performance of the TMD for damping a target linear resonance.

Considering now the second mode (Fig.~\ref{manifold_vari_abs}(b)), a large region, where the TBNES manifold lays above the TMD manifold, exists between point A$_2$ and D$_2$. This suggests that the TBNES might outperform the TMD in this specific conditions.

\begin{figure}
\begin{centering}
\setlength{\unitlength}{\textwidth}
\begin{picture}(1,0.33)
\put(0.07,0.0){\includegraphics[trim = 10mm 10mm 10mm 10mm,width=0.4\textwidth]{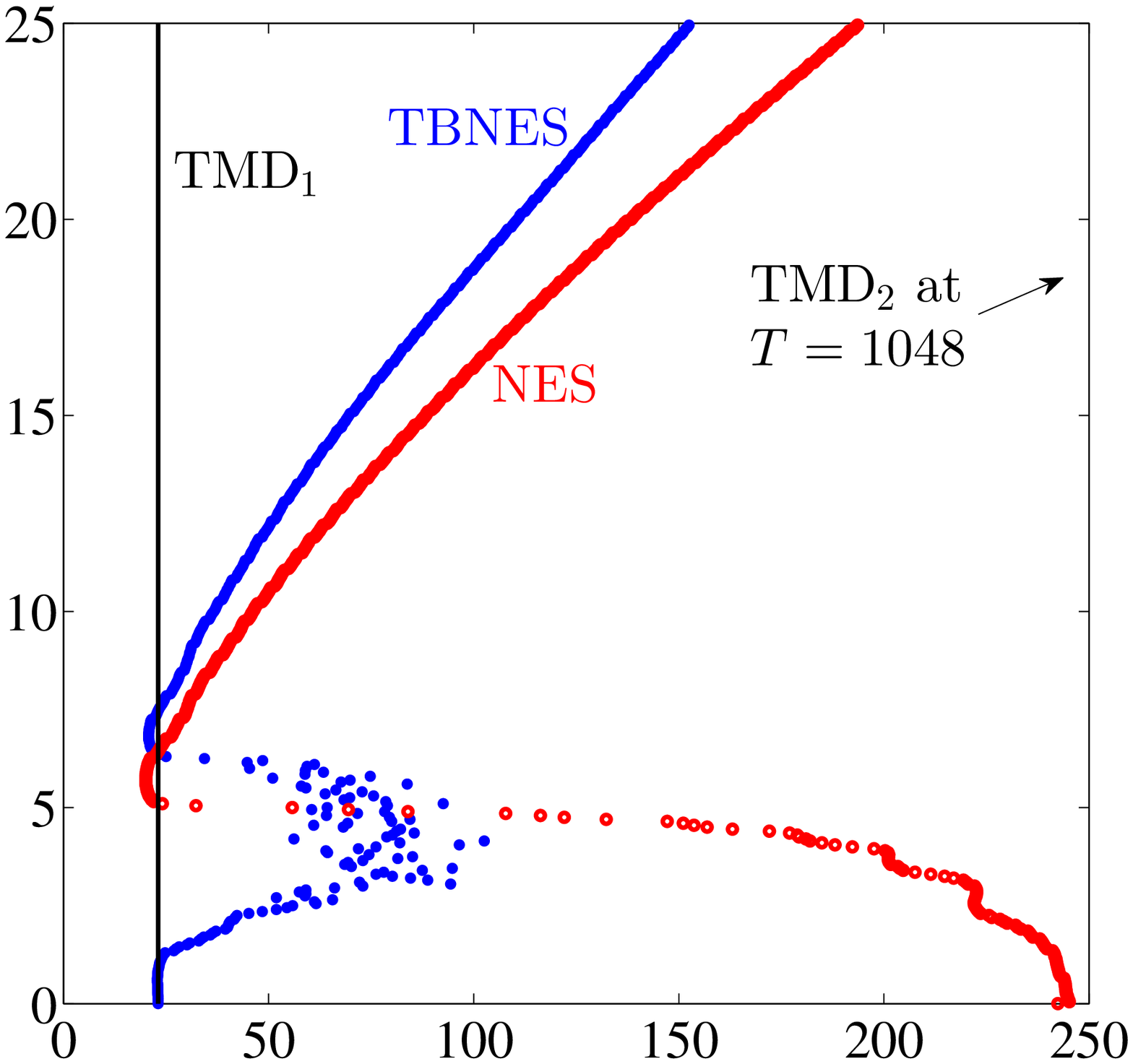}}
\put(0.51,0.0){\includegraphics[trim = 10mm 10mm 10mm 10mm,width=0.4\textwidth]{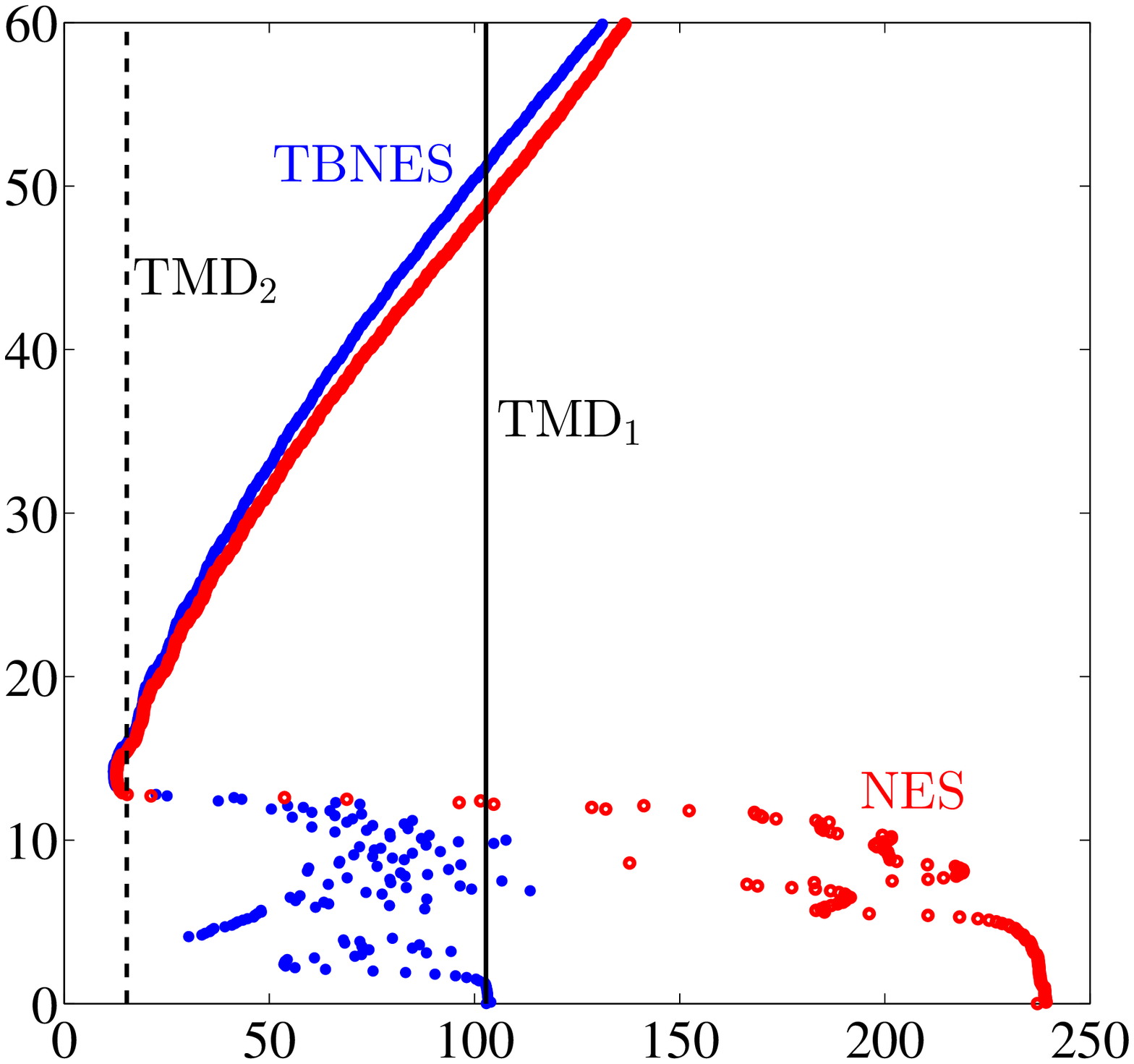}}
\put(0.05,0.33){(a)}
\put(0.49,0.33){(b)}
\put(0.04,0.19){$\dot y_1(0)$}
\put(0.48,0.19){$\dot y_2(0)$}
\put(0.275,-0.02){$T$}
\put(0.715,-0.02){$T$}
\par
\end{picture}
\end{centering}
\caption{\label{compar_time_1D}Dissipation time (70~\% of initial energy) for various initial conditions on the first (a) and on the second (b) mode. Blue dots: TBNES ($\lambda_3=0.005$, $\mu_2=0.1$, $\gamma=0.707$); red circles: NES ($\lambda_3=0.005$, $\mu_2=0.1$); black solid lines: TMD$_1$ (tuned on first mode) ($\gamma_{TMD}=1$, $\mu_2=0.1$); black dashed line: TMD$_2$ (tuned on second mode) ($\gamma_{TMD}=\sqrt 3$, $\mu_2=0.1$). Initial conditions $y_1=y_2=y_3=\dot y_3=0$ (except for TBNES $y_3=-\gamma/\sqrt{\lambda_3}$). $\varepsilon=0.05$ in all cases.}
\end{figure}

This occurrence is verified numerically, as illustrated in Fig.~\ref{compar_time_1D}.
Fig.~\ref{compar_time_1D}(a) depicts the time required to dissipate 70~\% of initial energy by the TBNES, the NES and the TMD (tuned either to the first or to the second mode) for initial conditions involving the first mode only.
The TMD (TMD$_1$ for the first mode) always outperform the TBNES. For very small initial energy the TBNES and the TMD are practically identical, as the linearized in-well dynamics of the TBNES correspond to the TMD.
For larger impulses, the performance of the TBNES worsens (while the TMD is amplitude invariant).
Interestingly, the tangency of the manifolds in Fig.~\ref{manifold_vari_abs}(a) correspond to an almost tangency of the numerical performance curves in  Fig.~\ref{compar_time_1D}(a).

Considering impulses on the second mode, Fig.~\ref{compar_time_1D}(b) clearly illustrates how the TBNES outperforms the TMD for a large range of initial energy.
Also in this case, for low amplitudes the TBNES and the TMD are equivalent, however, now the nonlinear characteristic of the TBNES improves its performance for increasing energy levels.
By tuning the TMD according to the second mode (TMD$_2$ in the figure), its performance on the second mode is naturally enhanced.
However, on the first mode, it is greatly deteriorated (dissipation time goes to 1048 time units) making the absorber practically useless.

Acknowledging the similar shape of the manifolds, and the slight influence of $\gamma$ for large amplitudes, we can argue that, at high energy level, the TBNES and the NES have analogous behavior.
This is verified numerically in Figs.~\ref{compar_time_1D}(a) and (b).
In both figures, for initial impulse above a critical value, the trend of the dissipation time is very similar and the differences between the two absorbers retrace the (minimal) differences depicted by the manifolds in Fig.~\ref{manifold_vari_abs}.
On the other hand, for low impulsive energy, the performance of the NES drastically decreases, which is due to the well-known low energy threshold required to activate the NES.
The TBNES, instead, does not present any minimal energy level, which is a significant advantage for practical applications.

\begin{figure}
\begin{centering}
\setlength{\unitlength}{\textwidth}
\begin{picture}(1,0.8)
\put(-0.07,0.8){\includegraphics[trim = 0mm 210mm 5mm 5mm,width=\textwidth]{Colorbar_BN.eps}}
\put(0.04,0.42){\includegraphics[trim = 10mm 10mm 10mm 10mm,width=0.4\textwidth]{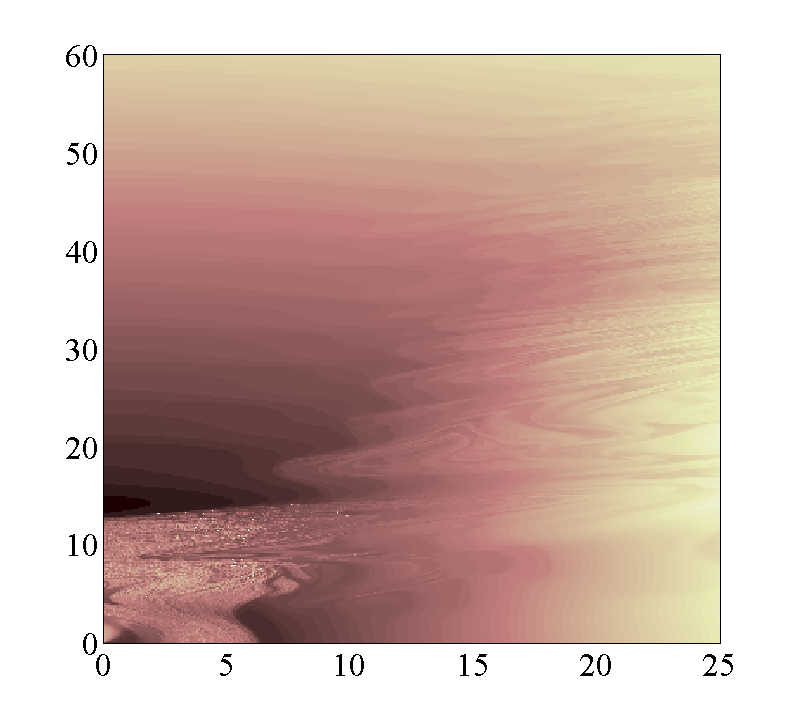}}
\put(0.46,0.42){\includegraphics[trim = 10mm 10mm 10mm 10mm,width=0.4\textwidth]{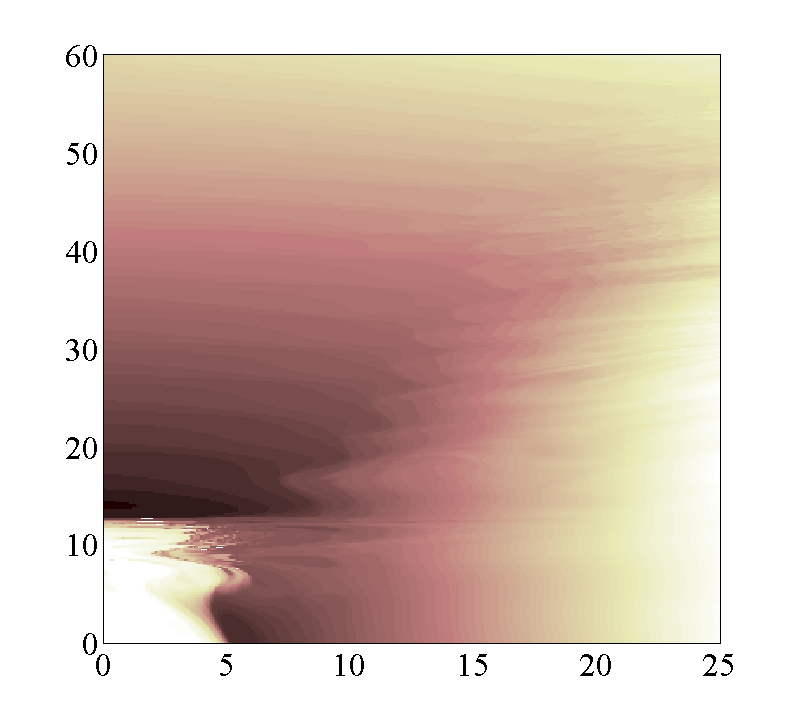}}
\put(0.04,0.02){\includegraphics[trim = 10mm 10mm 10mm 10mm,width=0.4\textwidth]{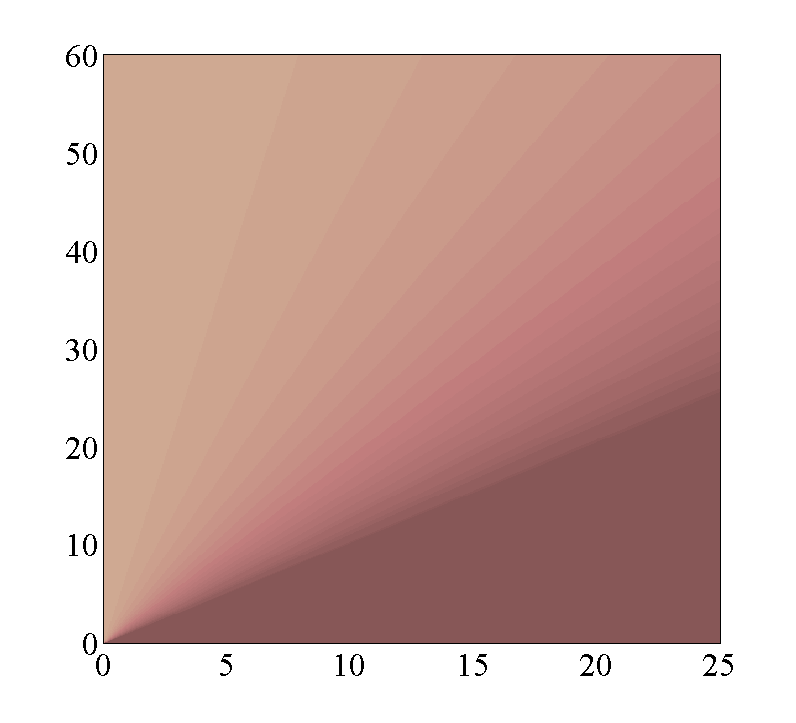}}
\put(0.46,0.02){\includegraphics[trim = 10mm 10mm 10mm 10mm,width=0.4\textwidth]{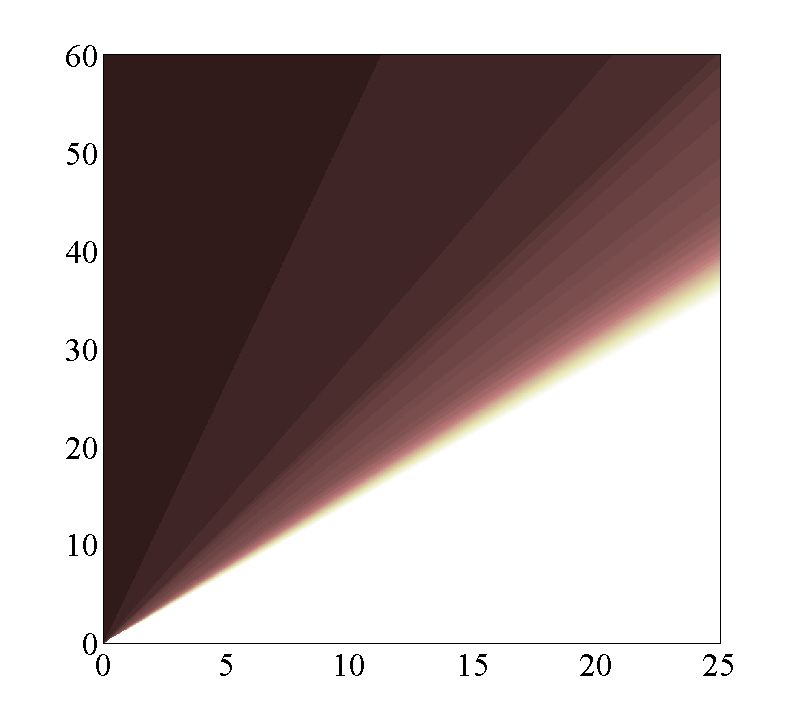}}
\put(0.02,0.81){$T$}
\put(0.02,0.76){(a)}
\put(0.44,0.76){(b)}
\put(0.02,0.36){(c)}
\put(0.44,0.36){(d)}
\put(0.01,0.61){$\dot y_2(0)$}
\put(0.43,0.61){$\dot y_2(0)$}
\put(0.01,0.21){$\dot y_2(0)$}
\put(0.43,0.21){$\dot y_2(0)$}
\put(0.23,0.4){$\dot y_1(0)$}
\put(0.65,0.4){$\dot y_1(0)$}
\put(0.23,0){$\dot y_1(0)$}
\put(0.65,0){$\dot y_1(0)$}
\par
\end{picture}
\end{centering}
\caption{\label{compar_time}Dissipation time (70~\% of initial energy) for various initial conditions and fixed absorber parameters. (a) TBNES $\lambda_3=0.005$, $\mu_2=0.1$, $\gamma=0.707$ (b) NES $\lambda_3=0.005$, $\mu_2=0.1$; (c) TMD (tuned on first mode) $\gamma_{TMD}=1$, $\mu_2=0.1$; (d) TMD (tuned on second mode) $\gamma_{TMD}=\sqrt 3$, $\mu_2=0.1$. Initial conditions $y_1=y_2=y_3=\dot y_3=0$ (except for TBNES $y_3=-\gamma/\sqrt{\lambda_3}$). $\varepsilon=0.05$ in all cases, simulations were limited to 200 time units.}
\end{figure}
The performance of the TBNES for impulses involving both modes is shown in Fig.~\ref{compar_time}(a), in terms of dissipation time.
The figure illustrates that the TBNES is able to work efficiently for a relatively wide range of energy levels, involving both modes. It thus works broadband both in amplitude and in frequency.
We notice that very high energy levels ($\dot y_1>20$ or $\dot y_2>50$) cause a deterioration of its performance.
Nevertheless, this limit can be conveniently adjusted through a tuning of $\lambda_3$.

Also in this case, the NES (Fig.~\ref{compar_time}(b)) behaves similarly to the TBNES for large amplitude.
However, as already discussed, a minimum energy threshold is needed to activate the NES, evidencing once more the superior performance of the TBNES.

Considering the same set of initial conditions, the TMD (Figs.~\ref{compar_time}(c) and (d)) works efficiently only for the targeted mode.
The figures illustrate on the one hand the advantage of the TBNES over the TMD in terms of frequency broadband, on the other hand the advantage of the TMD in terms of unlimited amplitude band for the targeted mode.
Furthermore, Figs.~\ref{compar_time}(c) and (d) confirm the known TMD design criterion of targeting the lowest natural frequency, which guarantees an overall better performance.
Thus, the same criterion shell be extended to the in-well tuning of the TBNES.

\begin{figure}
\begin{centering}
\setlength{\unitlength}{\textwidth}
\begin{picture}(1,1)
\put(0.02,0.7){\includegraphics[trim = 10mm 10mm 10mm 10mm,clip,scale=0.3]{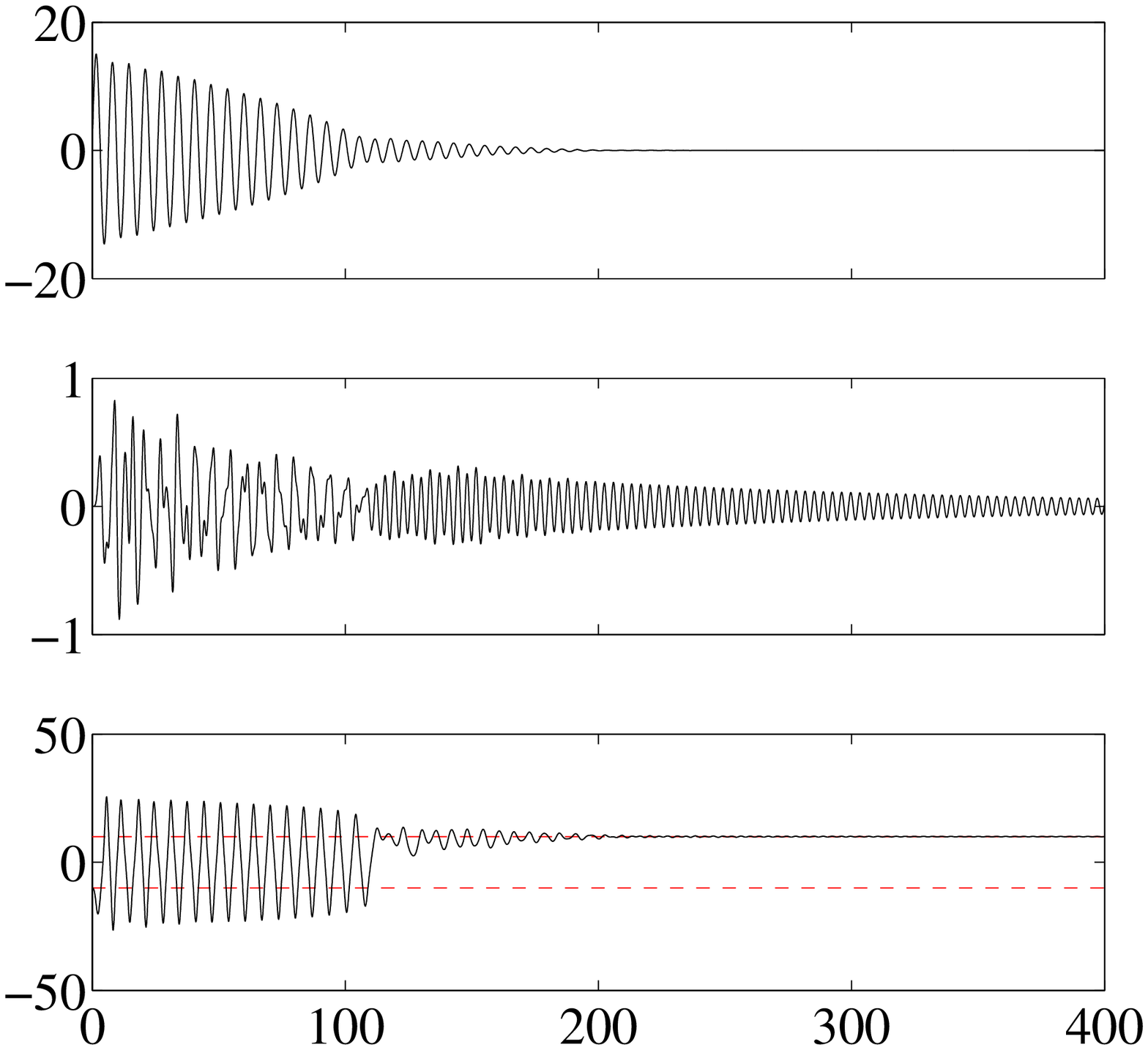}}
\put(0.36,0.7){\includegraphics[trim = 10mm 10mm 10mm 10mm,clip,scale=0.3]{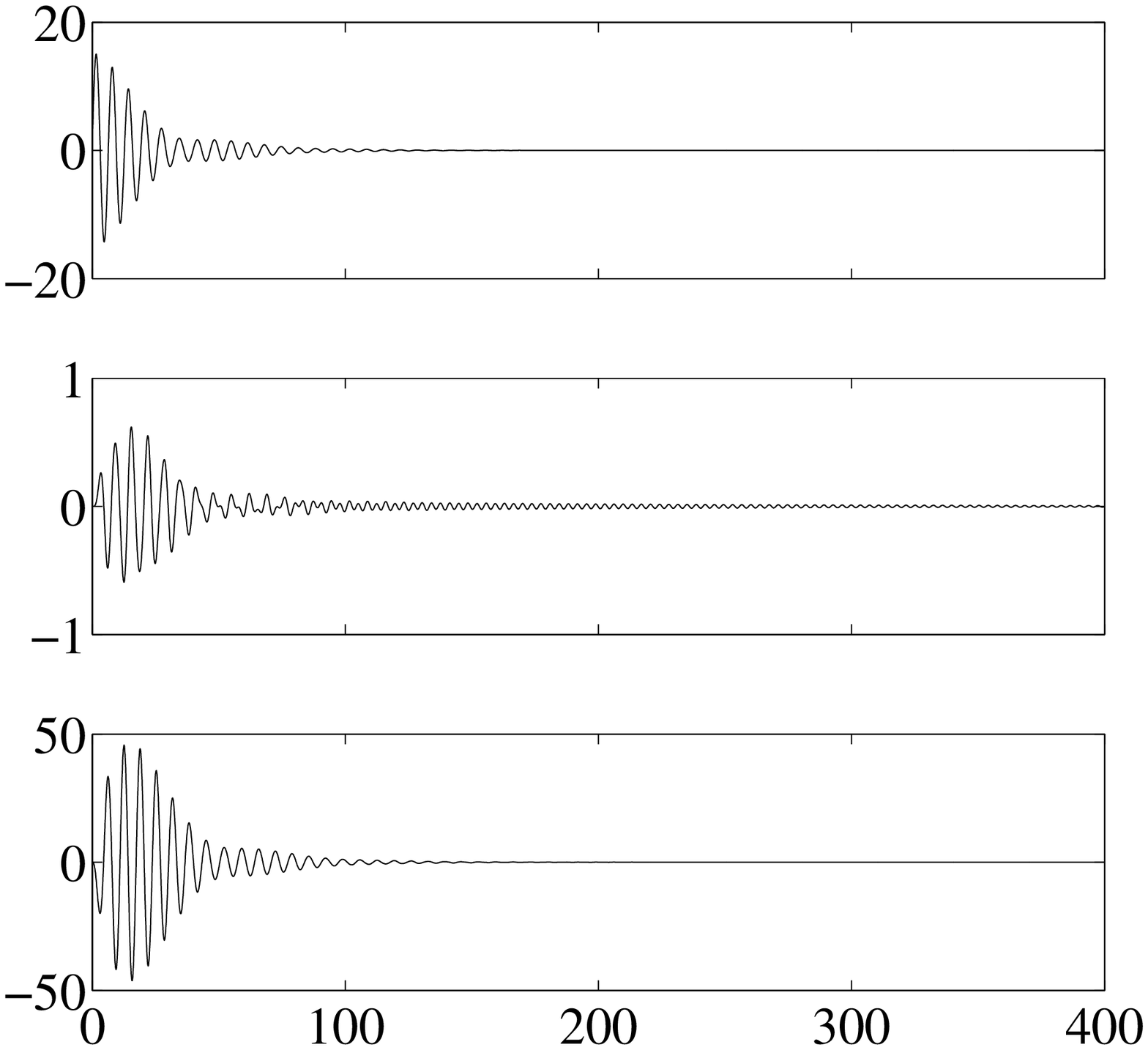}}
\put(0.70,0.7){\includegraphics[trim = 5mm 10mm 10mm 10mm,clip,scale=0.3]{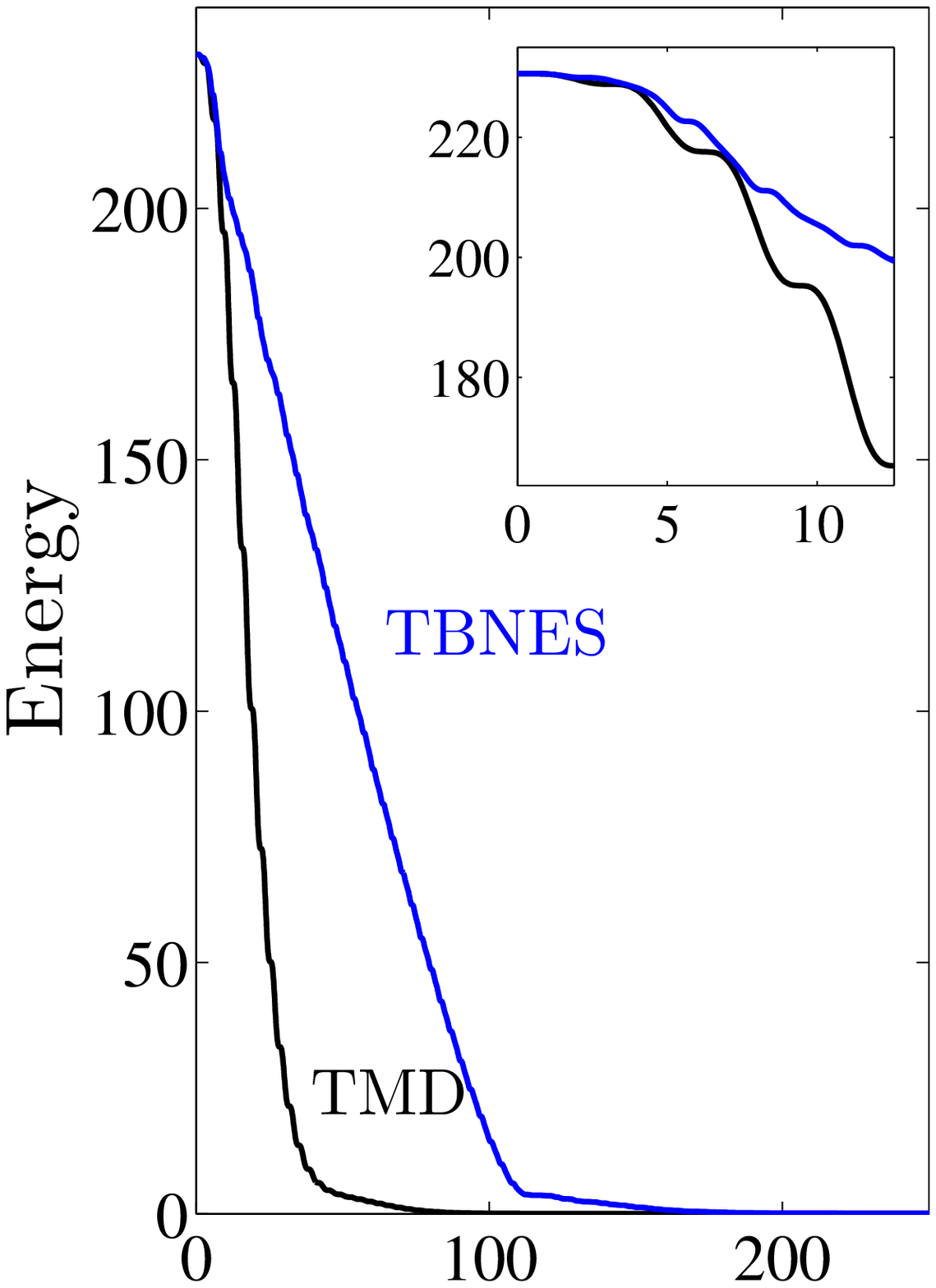}}
\put(0.02,0.35){\includegraphics[trim = 10mm 10mm 10mm 10mm,clip,scale=0.3]{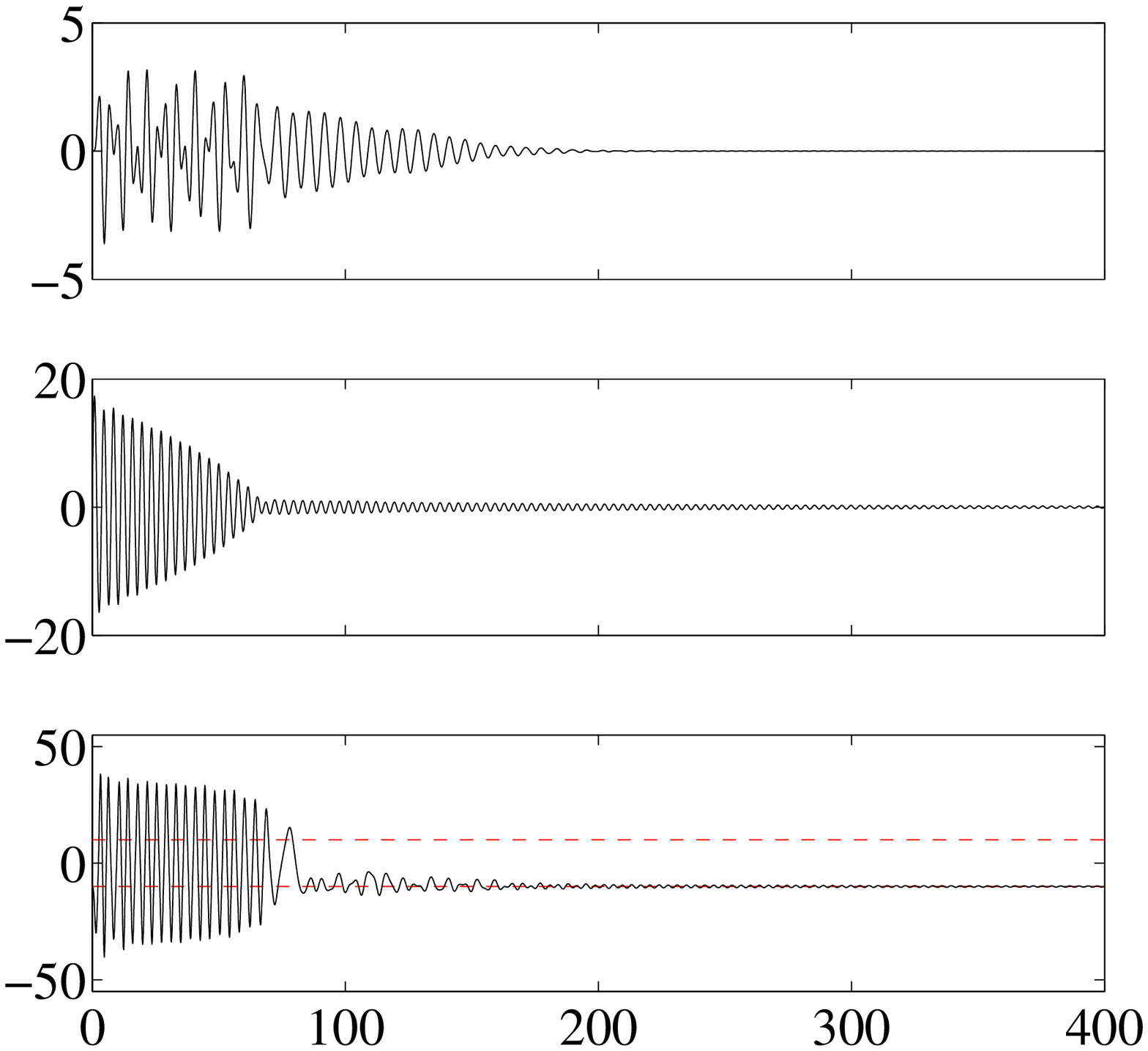}}
\put(0.36,0.35){\includegraphics[trim = 10mm 10mm 10mm 10mm,clip,scale=0.3]{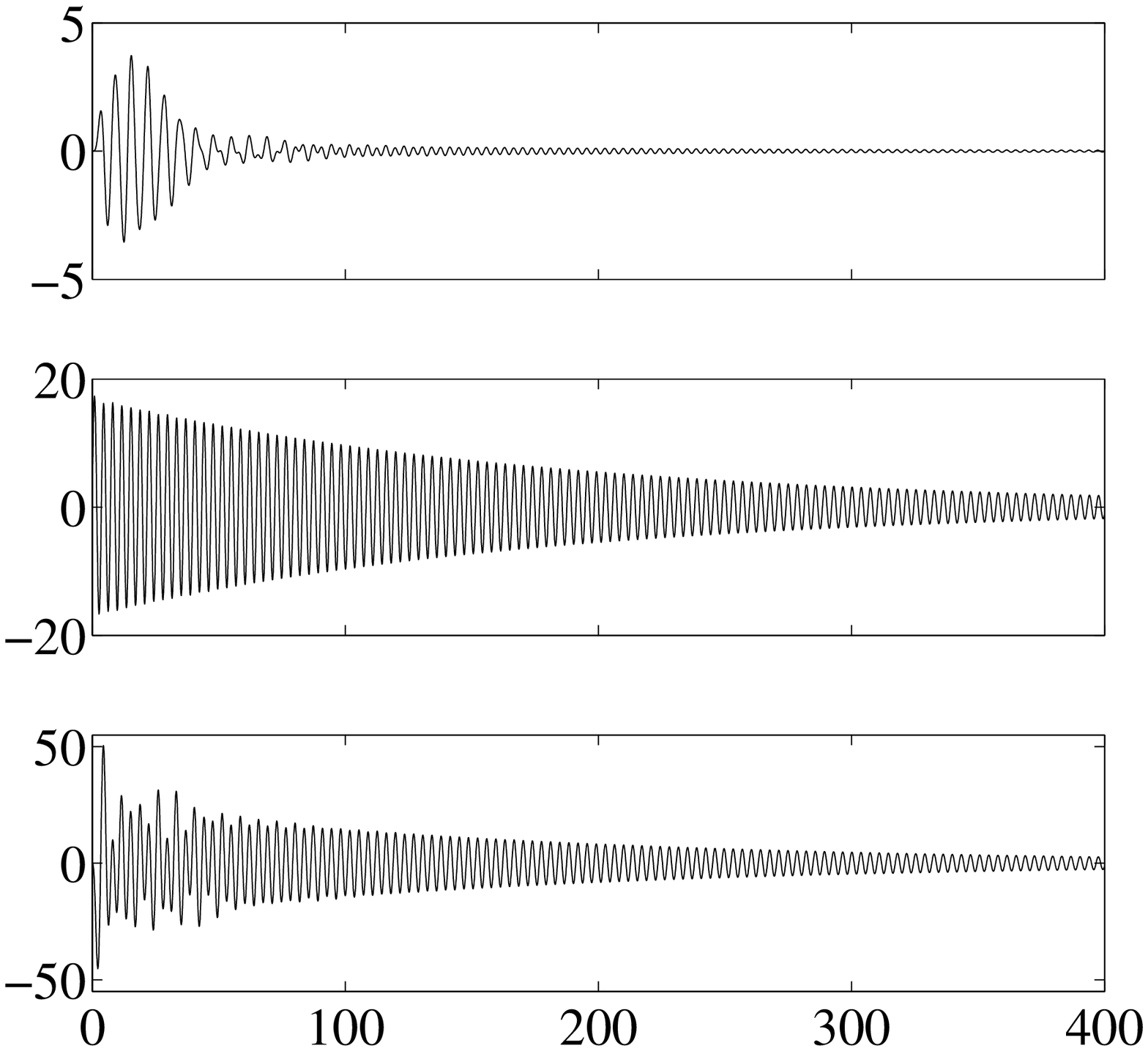}}
\put(0.70,0.35){\includegraphics[trim = 5mm 10mm 10mm 10mm,clip,scale=0.3]{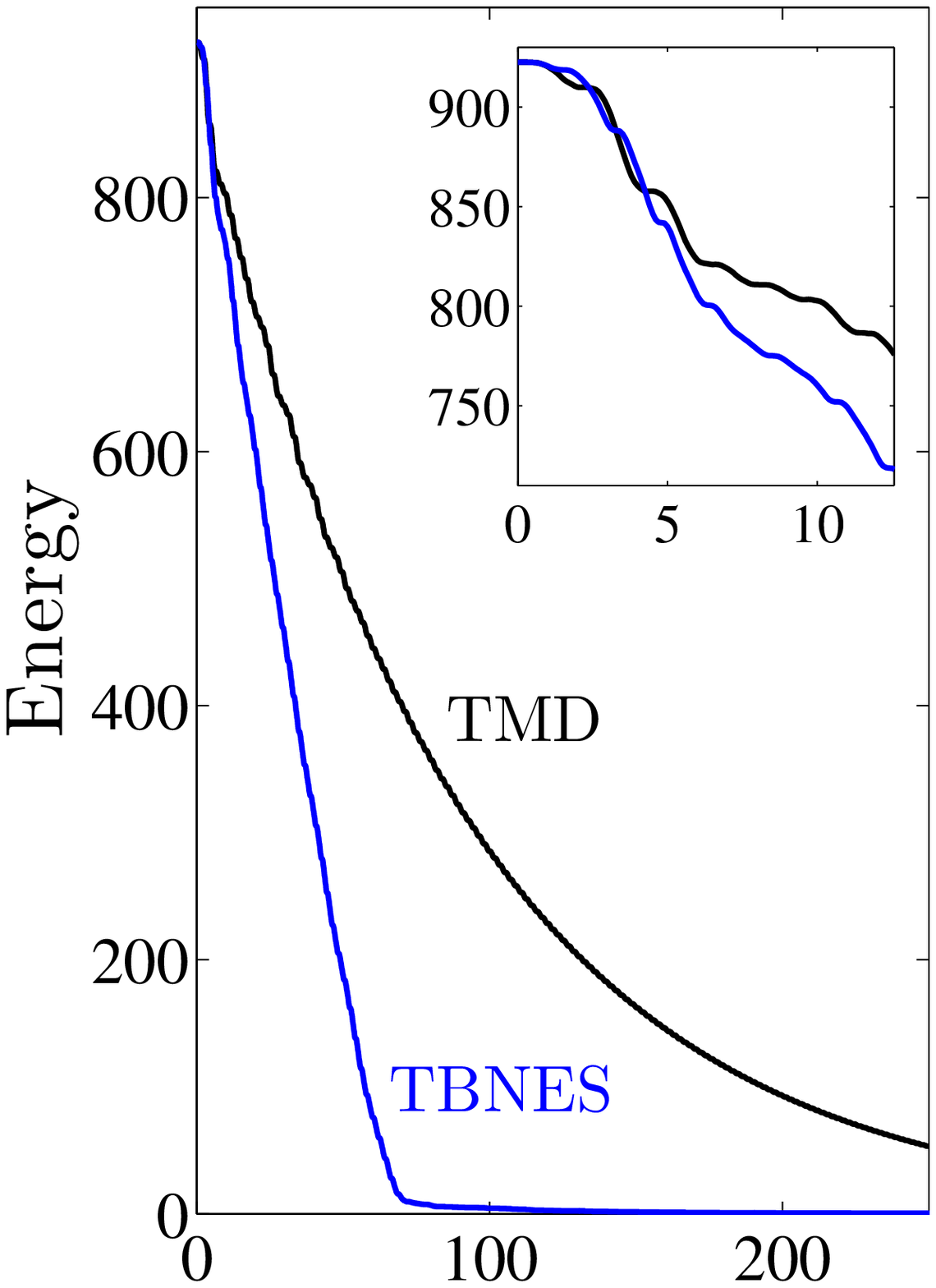}}
\put(0.02,0.0){\includegraphics[trim = 10mm 10mm 10mm 10mm,clip,scale=0.3]{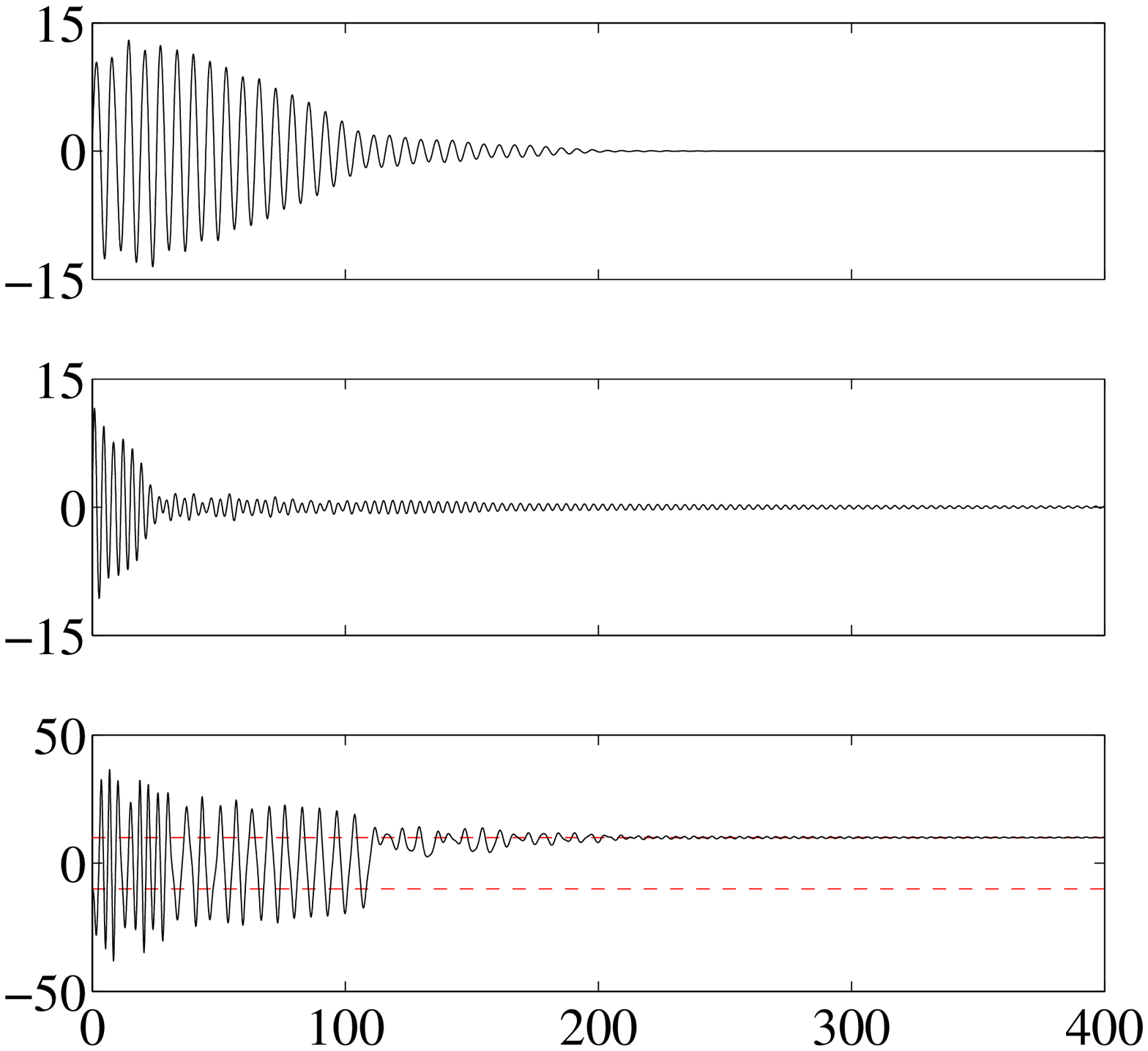}}
\put(0.36,0.){\includegraphics[trim = 10mm 10mm 10mm 10mm,clip,scale=0.3]{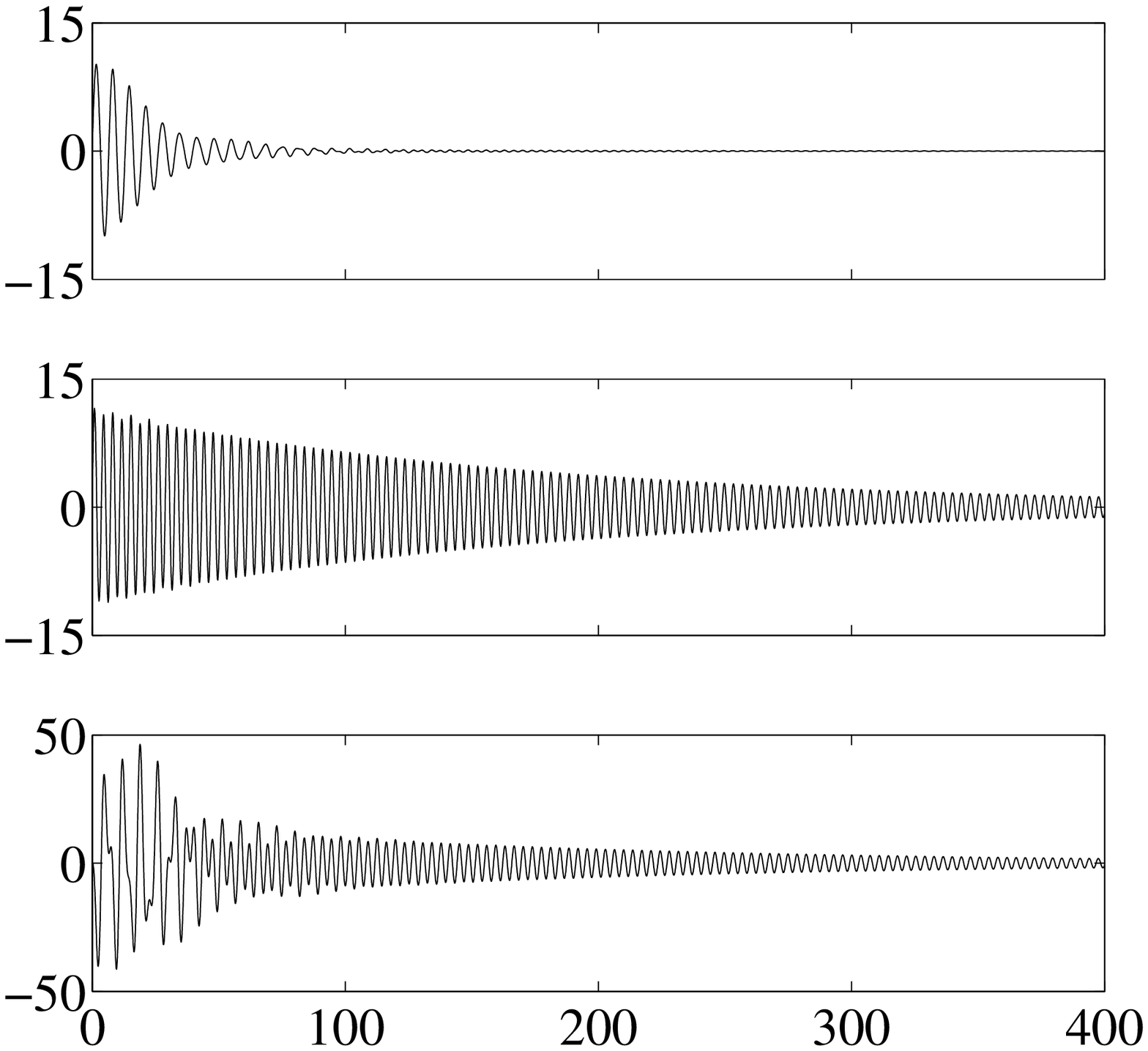}}
\put(0.70,0.){\includegraphics[trim = 5mm 10mm 10mm 10mm,clip,scale=0.3]{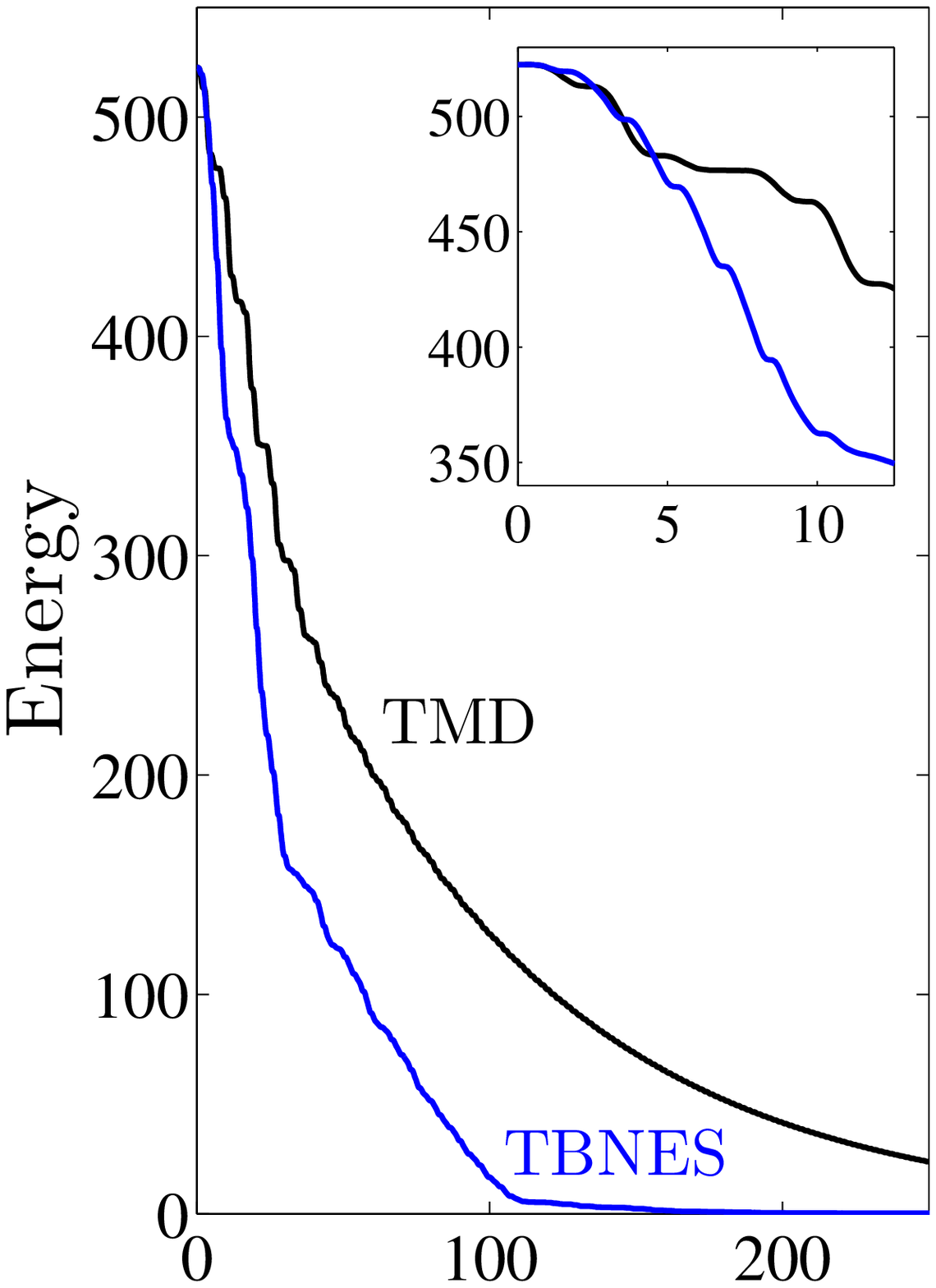}}
\put(0.0,0.99){(a)}
\put(0.34,0.99){(b)}
\put(0.70,0.99){(c)}
\put(0.0,0.64){(d)}
\put(0.34,0.64){(e)}
\put(0.70,0.64){(f)}
\put(0.0,0.29){(g)}
\put(0.34,0.29){(h)}
\put(0.70,0.29){(i)}
\put(0.01,0.96){$y_1$}
\put(0.01,0.86){$y_2$}
\put(0.01,0.76){$y_3$}
\put(0.01,0.61){$y_1$}
\put(0.01,0.51){$y_2$}
\put(0.01,0.41){$y_3$}
\put(0.01,0.26){$y_1$}
\put(0.01,0.16){$y_2$}
\put(0.01,0.06){$y_3$}
\put(0.35,0.96){$y_1$}
\put(0.35,0.86){$y_2$}
\put(0.35,0.76){$y_3$}
\put(0.35,0.61){$y_1$}
\put(0.35,0.51){$y_2$}
\put(0.35,0.41){$y_3$}
\put(0.35,0.26){$y_1$}
\put(0.35,0.16){$y_2$}
\put(0.35,0.06){$y_3$}
\put(0.19,0.68){$T$}
\put(0.53,0.68){$T$}
\put(0.83,0.68){$T$}
\put(0.19,0.33){$T$}
\put(0.53,0.33){$T$}
\put(0.83,0.33){$T$}
\put(0.19,-0.02){$T$}
\put(0.53,-0.02){$T$}
\put(0.83,-0.02){$T$}
\par
\end{picture}
\end{centering}
\caption{\label{time_series}Time series for TBNES (a,d,g) and TMD (b,e,h) and relative energy decrements (c,f,i). Parameter values $\lambda_3=0.005$, $\mu_2=0.1$, $\gamma=0.707$ (TBNES) and $\gamma_{TMD}=1$, $\mu_2=0.1$ (TMD). Initial conditions $y_1=y_2=\dot y_3=0$, $y_3=-\gamma/\sqrt{\lambda_3}$ for TBNES and $y_1=y_2=y_3=\dot y_3=0$ for TMD, while (a,b,c) $\dot y_1=15$, $\dot y_2=0$; (d,e,f) $\dot y_1=0$, $\dot y_2=30$; (g,h,i) $\dot y_1=10$, $\dot y_2=20$. $\varepsilon=0.05$ in all cases.}
\end{figure}
Figure~\ref{time_series} depicts several modal time-histories of the primary system with either a TBNES or a TMD; excitation on the first mode (a,b,c), on the second mode (d,e,f) and on both of them (g,h,i) are considered.
Once again, the better performance of the TMD, over the TBNES, for mitigating oscillations of the first mode can be recognized (Fig.~\ref{time_series}(c)).
On the contrary, the faster energy dissipation of the TBNES in the other two cases is also clearly illustrated (Figs.~\ref{time_series}(f)-(i)).
Time series clearly depicts the different performance of the TMD for mitigating motions of the targeted and of the non-targeted mode.
Regarding the TBNES, the various kinds of motions (large periodic intra-well and in-well) are well recognizable.
Instead, the chaotic regime cannot be easily identified, for the considered cases, and, apparently, it does not assume an important role for the dissipation. The situation might differ for other initial conditions.

The prompt responsiveness of a vibration absorber is another relevant parameter for its implementation in real applications.
In many real-life situations, the ability of the absorber to act immediately is more important, in terms of safety, than its capacity to rapidly damp small amplitude oscillations.
Enlargements of the energy decrements in Figs.~\ref{time_series}(c), (f) and (i) are depicted in their relevant insets, referring to the first two periods of oscillation.
The qualitative trend of the curves is analogous during the initial two periods and during the considered longer time frame (the TMD outperforms the TBNES in Fig.~\ref{time_series}(c), while the TBNES is more effective than the TMD in Figs.~\ref{time_series}(f) and (i)), this suggests that the efficient behavior of the TBNES discussed so far, reflects its operational responsiveness.

\section{Conclusions}

By adopting targeted analytical treatments for low and high amplitude dynamics, we have described a procedure to tune and optimize a bistable nonlinear energy sink (BNES) to obtain what we called a TBNES.
Having fixed a small mass ratio $\varepsilon$, the TBNES performance is ruled by three dimensionless parameters, the proportional damping $\mu_2$, the negative stiffness $\gamma$ and the cubic restoring force $\lambda_3$.
According to the reported findings, in order to optimize the in-well dynamics, $\gamma$ shell be tuned targeting the lowest natural frequency of the primary system. Moreover, $\lambda_3$ shell be chosen according to the amplitude range of interest, while $\mu_2$ shell be adjusted to smooth the transitions between the different dynamic regimes and to enlarge the range of efficiency both in amplitude and in frequency.
Encompassing two properties usually incompatible, the resulting TBNES is able to efficiently mitigate oscillations of the primary system at different energy levels and for more than one mode of vibration.
The absorption capabilities of the TBNES are superior to those of the NES.
Furthermore, by broadening the operational frequency range, the TBNES is able to overcome the TMD shortcomings.

\section*{Acknowledgement}
G. Habib would like to acknowledge the financial support of the Belgian National Science Foundation FRS-FNRS (PDR T.0007.15) and of the European Union, H2020 Marie Sk\l odowska-Curie Individual Fellowship, Grant Agreement 704133.

%%%%%%%%%%%%%%%%%%%%%%%%%%%%%%%%%%%%%%%%%%%%%%%%%%%%%%%%%%%%%

\bibliographystyle{ieeetr}
\bibliography{references}

\end{document}